%% file: eadiekellerharris2018_submit2018-10-04arxiv.tex
\documentclass[twocolumn, 11pt]{aastex61}

\usepackage{natbib}
\usepackage{graphicx}
\usepackage{bm}
\usepackage{amsmath}
\usepackage{amsfonts}
\usepackage{amssymb}
\newcommand{\unit}[1]{\ifmmode \:\mbox{\rm #1}\else \mbox{#1}\fi} % units in math mode in roman font
\newcommand{\mamo}[1]{\mbox{$#1$}}
\newcommand{\mone}{\mamo{^{-1}}}

\def\sun{\hbox{$\odot$}}

\newcommand{\kms}{\unit{km~s\mone}}

\newcommand{\kpc}{\unit{kpc}}

\newcommand{\msun}{\mamo{M_{\sun}}}

\newcommand\T{\rule{0pt}{2.6ex}}       % Top strut
\published{2018 Sep 24 in The Astrophysical Journal}
\journalinfo{Published in \apj ~ on 2018 September 24}
\shortauthors{Eadie, Keller, \& Harris}
\begin{document}
	
	\title{Estimating the Milky Way's mass via hierarchical Bayes: \\A blind test on MUGS2 simulated galaxies}

	\correspondingauthor{Dr. Gwendolyn M. Eadie}
	\email{eadieg@uw.edu}
		\author[0000-0003-3734-8177]{Gwendolyn M. Eadie}
	\affil{Department of Astronomy,	University of Washington, Seattle, WA, USA}
	\affil{eScience Institute, University of Washington, Seattle, WA, USA}
	\affil{DIRAC Institute, University of Washington, Seattle, WA, USA}
	
	\author[0000-0002-9642-7193]{Benjamin W. Keller}
	\email{benjamin.keller@uni-heidelberg.de}
	\affil{Astronomisches Rechen-Institut, 	Universität Heidelberg,	Heidelberg, Germany}
	
	\author[0000-0001-8762-5772]{William E. Harris}
	\email{harris@physics.mcmaster.ca}
	\affil{Department of Physics \& Astronomy, McMaster University, Hamilton, Canada}

\begin{abstract}
		
In a series of three papers, Eadie et al. (2015, 2016, 2017b) developed a hierarchical Bayesian method to estimate the Milky Way Galaxy's mass given a physical model for the potential, a measurement model, and kinematic data of test particles such as globular clusters (GCs) or halo stars in the Galaxy's halo. The Galaxy's virial mass was found to have a 95\% Bayesian credible region (c.r.) of $(0.67, 1.09) \times 10^{12}\msun$ (Eadie et al 2017a,b). In the present study, we test the hierarchical Bayesian method against simulated galaxies created in the McMaster Unbiased Galaxy Simulations 2 (MUGS2), for which the true mass is known. We estimate the masses of MUGS2 galaxies using GC analogues from the simulations as tracers. The analysis, completed as a blind test, recovers the true \textbf{$M_{200}$} of the MUGS2 galaxies within 95\% Bayesian c.r. in 8 out of 18 cases. Of the ten galaxy masses that were not recovered within the 95\% c.r., a large subset have posterior distributions that occupy extreme ends of the parameter space allowed by the priors. A few incorrect mass estimates are explained by the exceptional evolution history of the galaxies. We also find evidence that the model cannot describe both the galaxies' inner and outer structure simultaneously in some cases.  After removing the GC analogues associated with the galactic disks, the true masses were found more reliably (13 out of 18 were predicted within the c.r.). Finally, we discuss how representative the GC analogues are of the real GC population in the Milky Way.

\end{abstract}
	

\section{Introduction}\label{sec:intro}
		
	The total mass of the Milky Way (MW) Galaxy is not known within a factor of two 	\citep[see Figure~1 of ][ for a dramatic illustration of $M_{200}$ values]{wang2015}.
	This is both unfortunate and problematic, because our Galaxy's mass is a fundamental quantity in many areas of astrophysics, astronomy, and cosmology. To complicate matters, mass estimates tend to be reported differently, making comparisons difficult. For example, some studies report a mass within a specific distance from the Galactic center, whereas others report a virial mass or $M_{200}$ based on cosmological parameters.
		
	The wide range of mass estimates in the literature is the result of two major factors: method choice and data selection.	Inference methods take a variety of approaches, such as the timing argument \citep[eg][]{kahn1959ApJ}, the use of kinematic tracers to infer the gravitational potential \citep[e.g.][and many others]{little1987,wilkinsonevans1999,sakamoto2003,dehnen2006MNRAS, xue2008, lawmajewski2010ApJ, watkins2010,deasonetal2012MNRAS}, and more recently the direct comparison to cosmological simulations \citep[e.g][]{boylan2011MNRASmc,busha2011,patel2017MNRAS}. Different studies rely on different types of data and on different physical assumptions,  making disparities between results difficult to interpret.
	
	While each method has its own merits, the most popular approaches continue to use the kinematics of \emph{tracers} (e.g. globular clusters (GCs), halo stars, and stellar streams) to constrain the MW's gravitational potential, and thus its total mass. Since the \emph{Gaia} Data Release 2 on 25 April 2018,  a number of studies have already estimated the Galaxy's mass using GC and stellar dynamics from the \emph{Gaia} data \citep[e.g.][and Eadie 2018, in preparation]{Fritz2018arXiv, watkins2018arXiv, posti2018arXiv}.
	
	Using kinematic tracers has advantages, but it also presents challenges. Different types of tracers are available, and their kinematic data suffer from incomplete velocity measurements --- usually because only the line-of-sight velocity component is known. There are also differing degrees of measurement uncertainties. Incomplete velocity measurements hinder our understanding of the tracer population's velocity anisotropy, which has been shown to influence mass estimates. Moreover, it is common practice to include or exclude data based on their (in)completeness. Which type of tracer and what components of the data researchers choose to include or exclude may contribute to the overall uncertainty in the MW's mass.
	
	Thankfully, the situation is improving. Data from the \emph{Gaia} satellite \citep{perryman2001Gaia,GAIAweb} and the Large Synoptic Survey Telescope \citep[LSST;][]{LSSTweb} have and will greatly increase the number of kinematic tracers (e.g. RR Lyrae stars in the Galactic halo) and help overcome some challenges. The number of measurements for kinematic tracers will increase with these programs, especially with \emph{Gaia's} ability to measure proper motions and parallaxes of individual halo stars.
	
	With these ``big data'' come a demand for reliable methods that use tracer information to estimate the mass of the Galaxy. Moreover, methods with the potential to incorporate more than one type of tracer population are needed. A hierarchical Bayesian approach is ideal in this scenario, as tracer populations are following the same overall gravitational potential, but may have differing spatial distributions.
	
	In this vein, we have been developing a hierarchical Bayesian method to measure the mass and mass profile of the MW that uses kinematic tracers, called Galactic Mass Estimator (GME). GME has already been applied to the Galactic GC data \citep[][hereafter Papers 1, 2, and 3]{2015EHW,eadie2016, ESH2017ApJErratum, ESH2017}. In the last of these studies, GME provided a 95\% Bayesian credible region (c.r.) for the MW's virial mass: $(0.67,1.09) \times 10^{12} \msun$, which is in agreement with several recent studies
	 \citep[e.g.][]{xue2008,diaz2014MNRAS,gibbons2014,mcmillan2017MNRAS,patel2017MNRAS}. The median estimate for the virial mass of the MW was $0.86\times10^{12}\msun$.
	
	GME has at least three advantages over traditional mass estimation methods that use kinematic tracers: (1) incomplete and complete data are included simultaneously, (2) it uses a measurement model to account for observational uncertainty in position and velocity measurements, and (3) \textsc{GME} produces Bayesian c.r. for the \emph{cumulative} mass profile of the Galaxy at any galactocentric radius, rather than point estimates of the total mass within a certain distance. As shown in Paper 3, the cumulative mass profile with a Bayesian c.r. makes it easy to compare our results with estimates from other studies that report the mass within different distances from the Galactic center.
	
	In Papers 1--3, our method led to reasonable and encouraging results for the mass of the MW \citep[see also][]{eadiePhD2017}. As a next step, the hierarchical method could be extended for use with multiple tracer populations by adding another layer to the hierarchy. This is certainly a tempting avenue of research given the second Gaia data release in 2018. By allowing for different spatial distributions for each tracer population, and assuming they follow a parameterized model for the total gravitational potential of the Galaxy, we can hope to better estimate the mass of the MW.

	Before moving forward, however, it is important to address any uncertainty associated with the GME method thus far. Foremost, it remains unclear if the derived quantities from the posterior distribution correctly describe the \emph{true} total mass and cumulative mass profile of the Galaxy. In other words, we need to have a sense of how well our mass profile prediction represents the truth within the statistical uncertainties. We also want to better understand the limitations of the physical model (Section~\ref{sec:model4}), and to be able to recognize when the model has gone awry in light of the data.
	
	Therefore, a natural step is to test the hierarchical Bayesian method on mock observations derived from hydrodynamical simulations of MW-type galaxies, in order to obtain insight into the predictive properties of our choice of model. 
		
	In this study, we perform blind tests on simulated observations of GC analogues within galaxies created by the McMaster Unbiased Galaxy Simulations 2 (MUGS2) project  \citep[see][and Section~\ref{sec:MUGS2}]{keller2015,keller2016}. These hydrodynamical simulations incorporate the modern smoothed particle hydrodynamics code GASOLINE2 \citep{wadsley2004gasoline,wadsley2017gasoline2}, and include comprehensive effects such as low-temperature metal cooling \citep{shen2010MNRAS}, UV background radiation, star formation, and stellar and superbubble feedback \citep{keller2014MNRAS,keller2015}. 

	The mock galaxies provide a way to test our method's predictive power because their stellar and dark matter profiles are more complex than the physical model assumed by GME --- a similar situation when we apply our method to the real MW data.
		
\begin{figure*}
		\label{fig:galimages}

		\includegraphics[width=\linewidth]{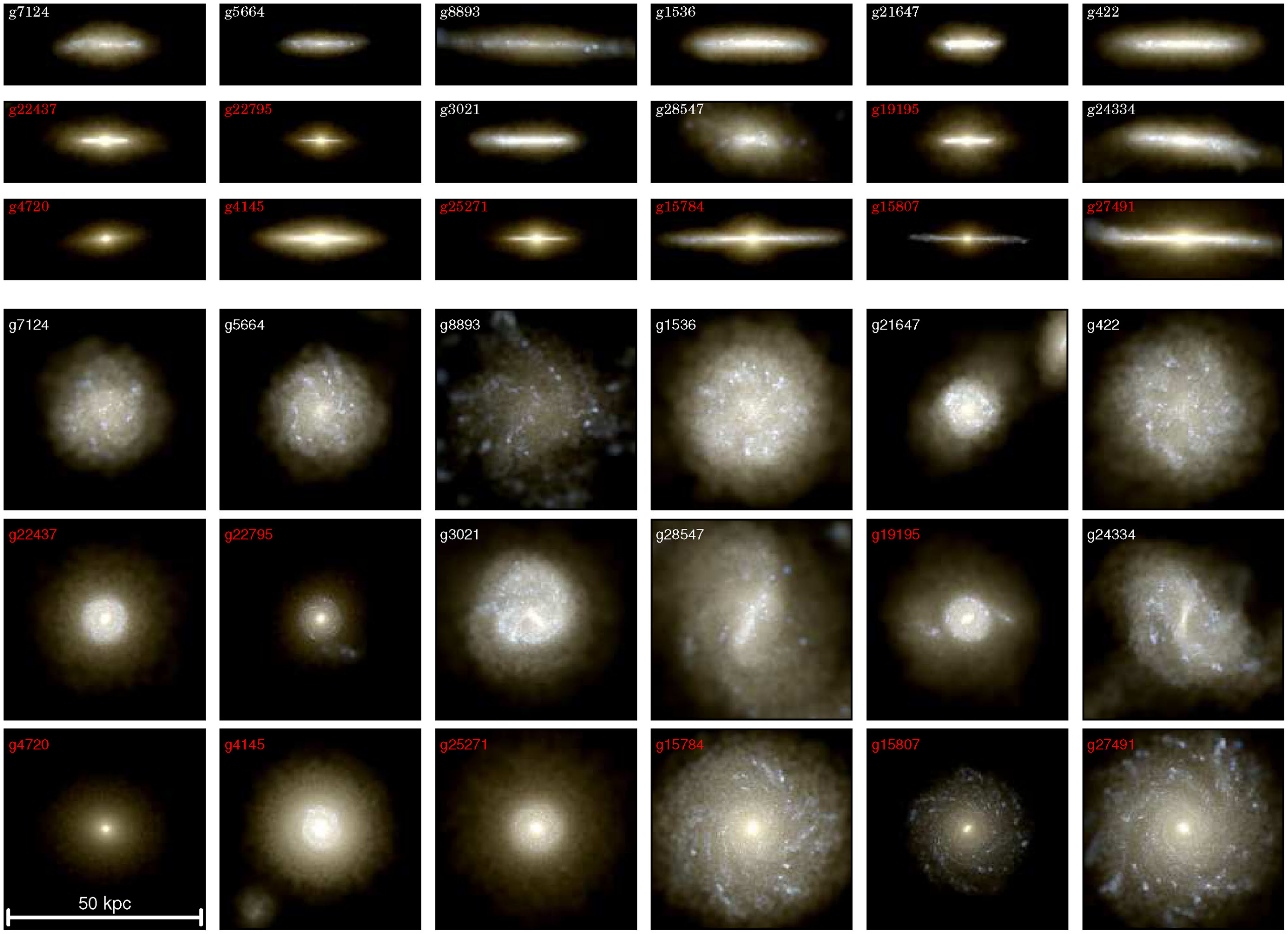}
		
		\caption{Mock images of the MUGS2 galaxies edge-on (top) and face-on (bottom). Reproduced from Figure 1 in \cite{keller2016}, ``Cosmological galaxy evolution with superbuble feedback - II. The limits of supernovae'', MNRAS, 453, 3499. Red labels denote the \emph{unregulated} galaxies introduced in Section~\ref{sec:MUGS2}.}

	\end{figure*}

	Eighteen galaxies were created by MUGS2, and we analyse each of them individually. Mock images of the galaxies are shown in Figure~\ref{fig:galimages} \citep[reproduced from][Figure 1]{keller2016}. We subject all of these galaxies to the blind test, and present detailed results for two galaxies ($g15784$ and $g1536$) as examples. Summarized results for the other galaxies are also provided, and used to make inferences about our method.
	
	The organization of this paper is as follows. Section~\ref{sec:MUGS2} provides a summary of the MUGS2 simulations that were carried out in previous studies \citep{keller2014MNRAS}. Next,  Section~\ref{sec:model4} briefly reviews the physical model and the important points of the hierarchical Bayesian framework. Section~\ref{sec:methods4} describes how mock tracers from the MUGS2 simulations were selected, and how mock heliocentric observations (with errors) were created from these tracers. In Section~\ref{sec:results4}, we show the results and cumulative mass profile predictions of $g15784$ and $g1536$, and compare these to the true quantities (which were revealed only after our analysis was complete). The results from all eighteen blind tests and a discussion follows in the same section. Section~\ref{sec:conclude4} provides a summary of our findings and avenues of future work.
	
	\section{Summary of MUGS2 Simulations}\label{sec:MUGS2}
	
	The McMaster Unbiased Galaxy Simulations 2 (MUGS2) \citep{keller2016} is a cosmological re-simulation
		of a sample of Milky-Way like halos originally presented in \citet{stinson2010MNRAS} (i.e. MUGS). The initial MUGS sample was not designed to specifically create a single object like the Milky Way, but rather to provide an unbiased sample of the kind of galaxy that lives in halos with masses	$\sim10^{12}\msun$, which is where we expect $L*$ galaxies to reside \citep{moster2010}. Both the MUGS and MUGS2 simulations use a \emph{Wilkinson
		Microwave Anisotropy Probe 3} $\Lambda$CDM cosmology with $H_0=73\kms\mp^{-1}$, $\sigma_8=0.76$, and components $\Omega_m=0.24$,
		$\Omega_\Lambda=0.76$ and $\Omega_{\text{bary}}=0.04$ \citep{spergel2007}.
	
	The MUGS2 simulations differed from the original \citet{stinson2010MNRAS} simulations in two major ways. First, while \citet{stinson2010MNRAS} was simulated with the hydrodynamics code GASOLINE \citep{wadsley2004gasoline}, MUGS2 was simulated with GASOLINE2 \citep{wadsley2017gasoline2}.  The latter includes a new subgrid model for turbulent mixing of metals and energy \citep{shen2010MNRAS}, and improved hydrodynamics \citep[see][for details]{wadsley2017gasoline2}. Second, MUGS2 introduced a physically motivated ``superbubble''
		model for stellar feedback; the new model captures the unresolved mixing between the cold, swept up shell and the hot interior of superbubbles driven by supernovae from
		star clusters \citep{keller2014MNRAS}.
	
	MUGS2 shared the same initial conditions as the original MUGS study, in order to investigate the effects of including improved hydrodynamic methods and a more realistic model for stellar feedback. The MUGS initial conditions were drawn from a set of 18 cosmological zoom-in
		\citep{quinn1992} galaxies, selected from a $50\;h^{-1}$Mpc box evolved with
		$256^3$ dark matter particles to $z=0$.  Halos from the simulation were selected based on their mass and isolation alone. For halos between $5\times10^{11}\msun$ and
		$2\times10^{12}\msun$, 267 halos had no similarly sized neighbours within less than
		$2.7$Mpc, and of these halos, 18 were selected randomly in order to sample the spin parameter and merger history space in an unbiased way. The 18 halos were then re-simulated at higher resolution, and particles that accreted within
		$3r_{\text{vir}}$ at $z=0$ were seeded with gas particles to generate a set of hydrodynamic initial conditions.
	
	The final sample of 18 galaxies have spin parameters $\lambda' = J/\sqrt{5/eGRM^3}$ between 0.009 and 0.106.  Their last major mergers occur over a range of redshifts, with $z=7.3$ for the most quiescent of the sample ($g15784$), to just before redshift 0 ($z=0.1$) for g28547.  The earliest galaxy to assemble half of its mass is g15784 ($z=1.3$), and the latest is g21647 ($z=0.2$).
	
	It was also found that the new feedback model in MUGS2 drives realistic,	mass-loaded winds from galaxies, and becomes ineffective at regulating star formation at the peak of the stellar-mass-to-halo mass curve	\citep{moster2010,keller2016} at $10^{12}\msun$.  Thus, those galaxies with halo	masses above this value contain 2-3 times too many stars, and significantly
		over-massive bulges.  In nature, feedback from active galactic nuclei (AGN)
		would become the dominant feedback process in these galaxies, but the MUGS2
		simulations omit AGN feedback. Therefore, the galaxies with halo masses
		$>10^{12}\msun$ are referred to as ``unregulated'', and the rest are referred to as ``regulated''.  The unregulated galaxies are distinguished by red labels in Figure~\ref{fig:galimages} (from \cite{keller2016}).
	
	During the blind tests of our method on the MUGS2 galaxies, we were not made aware of the unregulated and regulated categories. Thus, all galaxies were analyzed in the same way. Table~\ref{tab:numberoftracers} summarizes the physical characteristics of the MUGS2 galaxies, first grouped by regulated (upper-half) and unregulated (lower-half), and then listed in increasing mass.

	\input{MUGS2_results.tex}

	\section{Brief Review of the Hierarchical Bayesian Model}\label{sec:model4}
	
	We adopt the same physical model that was used in Papers 2 and 3 \citep[i.e. the model described by][]{evans1997, deason2011}. The model assumes a total gravitational potential given by
	\begin{equation}\label{eq:potential4}
	\Phi(r) = \frac{\Phi_o}{r^\gamma}
	\end{equation}
	where $\Phi_o$ and $\gamma$ are free parameters. The radial distribution of the tracer population is also assumed to follow a power-law profile:
	\begin{equation}\label{eq:tracers}
	\rho(r) \propto \frac{1}{r^{\alpha}},
	\end{equation}
	where $\alpha$ is a parameter. Given Equation~\ref{eq:potential4}, the total mass profile is given by
	\begin{equation}\label{eq:Mr4}
	M(r) = \frac{\gamma\Phi_o}{G}\left(\frac{r}{\text{kpc}}\right)^{1-\gamma},
	\end{equation}
	which goes to an isothermal sphere in the limit that $\gamma \rightarrow 0$, and a point mass as $\gamma \rightarrow 1$. Equations~\ref{eq:potential4} and \ref{eq:tracers} are used in the Eddington formula to derive a distribution function (DF) \citep[see][]{binney2008},
		\begin{equation}\label{eq:DFLfinal}
	f(\mathcal{E},L) = \frac{ L^{-2\beta}\mathcal{E}^{ \frac{\beta(\gamma-2)}{\gamma} + \frac{\alpha}{\gamma}-\frac{3}{2}} } {\sqrt{ 8\pi^{3} 2^{-2\beta}} \Phi_o^{-\frac{2\beta}{\gamma} + \frac{\alpha}{\gamma}}} \frac{
		\Gamma\left( \frac{\alpha}{\gamma} - \frac{2\beta}{\gamma}+ 1\right)}
	{\Gamma\left( \frac{\beta(\gamma-2)}{\gamma} + \frac{\alpha}{\gamma} -\frac{1}{2}\right)},
	\end{equation}
	where the model parameters are $\Phi_o, \gamma, \alpha$, and $\beta$. $\beta$ is the velocity anisotropy parameter (assumed constant) for the tracer population. For the curious reader, the derivation of the DF is given in both the original paper by \cite{evans1997}, albeit with different notation, and in Paper 2.	
	
	The DF in Equation~\ref{eq:DFLfinal} is a probability distribution; it gives the probability of  tracer $i$ having a specific energy $\mathcal{E}_i = -v_i^2/2 + \Phi(r_i)$,  and specific angular momentum $L_i = r_i v_{t_i}$, given the model parameters. The data are the total speed of the tracer $v_i = \sqrt{v^2_{i,r}+ v^2_{i,t}}$, and its distance from the Galactic centre $r_i = \sqrt{x^2_i + y^2_i + z^2_i}$. The velocity components $v_{i,r}$ and $v_{i,t}$ are the galactocentric radial and tangential velocity components. Assuming that the individual tracers in the population are independent, then the probability that all tracers have $\{\mathcal{E}_i, L_i\}$ is
	\begin{equation}
	\prod_{i=1}^{N} f(\mathcal{E}_i, L_i).
	\end{equation}
	
	The DF $f(\mathcal{E}, L)$ assumes a Galactocentric reference frame, but the data are measured from our heliocentric perspective. Transforming from one frame to the other is not difficult, but properly propagating the measurement uncertainties to the Galactocentric frame is. To overcome complex error propagation, we use a measurement model at the data level of the Bayesian hierarchy. That is, the measurement model is the likelihood in our hierarchical Bayesian analysis. 
	
	The measurement model (outlined more fully in Paper 3) assumes that a measurement of a quantity $x$ (e.g. line-of-sight velocity) is a random variable $X$ that is normally distributed about mean $\mu$,
	\begin{equation}\label{eq:normal}
	X \sim \mathcal{N}(\mu, \sigma^2),
	\end{equation}
 where the variance $\sigma^2$ is set equal to the square of the known measurement uncertainty. In other words, the data (position $r$ and velocity components $v_{los}$, $\mu_{\alpha} \cos{\delta}$, and $\mu_{\delta}$) are assumed to be drawn from normal distributions centered on the true but unknown position and velocity components in the Heliocentric frame. The true position and velocity components of each tracer are free parameters.
	
	The prior on the likelihood is the DF (Equation~\ref{eq:DFLfinal}). The positions and velocities of the tracers are assumed to have the spatial distribution given by Equation~\ref{eq:tracers} and to follow the influence of the total gravitational potential (Equation~\ref{eq:potential4}). In this way, each tracer has individual parameters for their true position and velocity, but also shares with the rest of the tracers the $\Phi_o$ and $\gamma$ parameters defining the total gravitational potential. 
	 
	The four model parameters $\Phi_o, \gamma, \alpha$, and $\beta$ in the DF are assigned hyperprior probability distributions, whose forms are given in Paper 3 and repeated here in Table~\ref{tab:priors}. We choose truncated, uniform prior probabilities for $\Phi_o, \gamma$, and $\beta$, and a Gamma distribution for the prior on $\alpha$ (see Papers 2 and 3 for justification and more details). There is no direct prior on the mass ($M_{200}$) since this quantity is determined by $\Phi_o$ and $\gamma$ through Equation~\ref{eq:Mr4}.  
	 \\
	 % table showing prior distributions
	 \begin{table}[t]
	 	\centering
	 	\caption{Hyperprior Probability Distributions}
	 	\label{tab:priors}
	 	\begin{tabular}{ccc}
	 		\tableline

	 		Parameter & Distribution & Hyperparameters \\
	 		\hline
	 		\hline
	 		$\Phi_o$& Uniform & $\Phi_{o,\text{min}}=1$, $\Phi_{o,\text{max}}=200$ \\  
	 		$\gamma$& Uniform  &$\gamma_{\text{min}}=0.3$, $\gamma_{\text{max}}=0.7$ \\
	 		$\alpha$ & Gamma & $b=0.4\kpc$, $c=0.001$, $p=0.001$ \\
	 		$\beta$ & Uniform &  $\beta_{\text{min}}=-0.5$, $\beta_{\text{max}}=1$ \\
	 		\tableline
	 	\end{tabular}
	 \end{table}
 
	In summary, the posterior distribution is given by 
	\begin{equation}
		\text{Posterior}\propto \text{ Likelihood } \times \text{Prior} \times \text{Hyperpriors}		
	\end{equation}
	\citep[see][]{eadieIAU2016, ESH2017ApJErratum, ESH2017,eadiePhD2017}. In standard Bayesian inference, the posterior distribution contains all the information about the model parameters, given the data, the model, and the prior information.
	
	The posterior distribution is most often approximated by generating a Markov chain that is a stationary distribution proportional to the posterior. We use this approach, but with the variation of a hybrid-Gibbs within the Metropolis sampler to increase efficiency. For information on the particulars of the Markov Chain Monte Carlo (MCMC) sampling methods and convergence diagnostics, the reader may refer to Paper 1, with some updates in Papers 2 and 3.
		
	In Paper 3, we estimated the virial mass and cumulative mass profile of the MW using GME and the GC population. We now apply this method to mock observations from the MUGS2 galaxies' simulated tracer data via blind tests.

	\section{Creating Mock Observations}\label{sec:methods4}
	We used GCs as tracers in our analysis of the MW. In order for our blind test to be as realistic as possible, we need tracer data from the simulated galaxies that most closely resemble that of the MW's GC population. In the next few sections, we describe the process for generating such mock data.
		
	\subsection{Finding GC analogues in MUGS2 galaxies}\label{sec:GCanalogs}
	Great strides in cosmological, hydrodynamical simulations have been made in recent years, but resolution limits have prevented the ability to create GC populations within a fully simulated, cosmological galactic environment\footnote{with the recent and notable exception of \cite{pfeffer2017}}. Instead, we must use a selection of ``star particles'' from each mock galaxy and treat them as \emph{GC analogues}. The star particles represent entire populations of stars, with each particle carrying a mass of $\sim 10^5 \msun$. Coincidentally, this mass is similar in mass to many GCs. 
	
	The GC analogues from the MUGS2 data are star particles with ages greater than 12 billion years and metallicities $[Fe/H] < -1.5$. With such cuts, the stars in a GC analogue would have formed at an approximate redshift $z\approx 3$ or higher. Disk-associated objects were also excluded by removing star particles within a galaxy-centered cylinder with radius $3r_{e}$ and height $r_{e}$, where $r_{e}$ is the half-light radius of the galaxy.
	
	In order to keep our analysis a true blind test,  only the galactocentric positions $(x,y,z)$ and velocities $(v_x, v_y, v_z)$  of the GC analogues were made known to us. The GC analogues may be bound or unbound from the host galaxy, but this information was withheld until the Bayesian solution was complete. This is an important point, as our hierarchical Bayesian method assumes all tracers are bound. Aside from the kinematic information and total number of GC analogues, and the knowledge that the MUGS2 host galaxies were ``Milky Way-type'' galaxies, we had no knowledge of their mass, mass profile, or merger history.
	
\begin{figure*}
	\centering
	\includegraphics[scale=0.54]{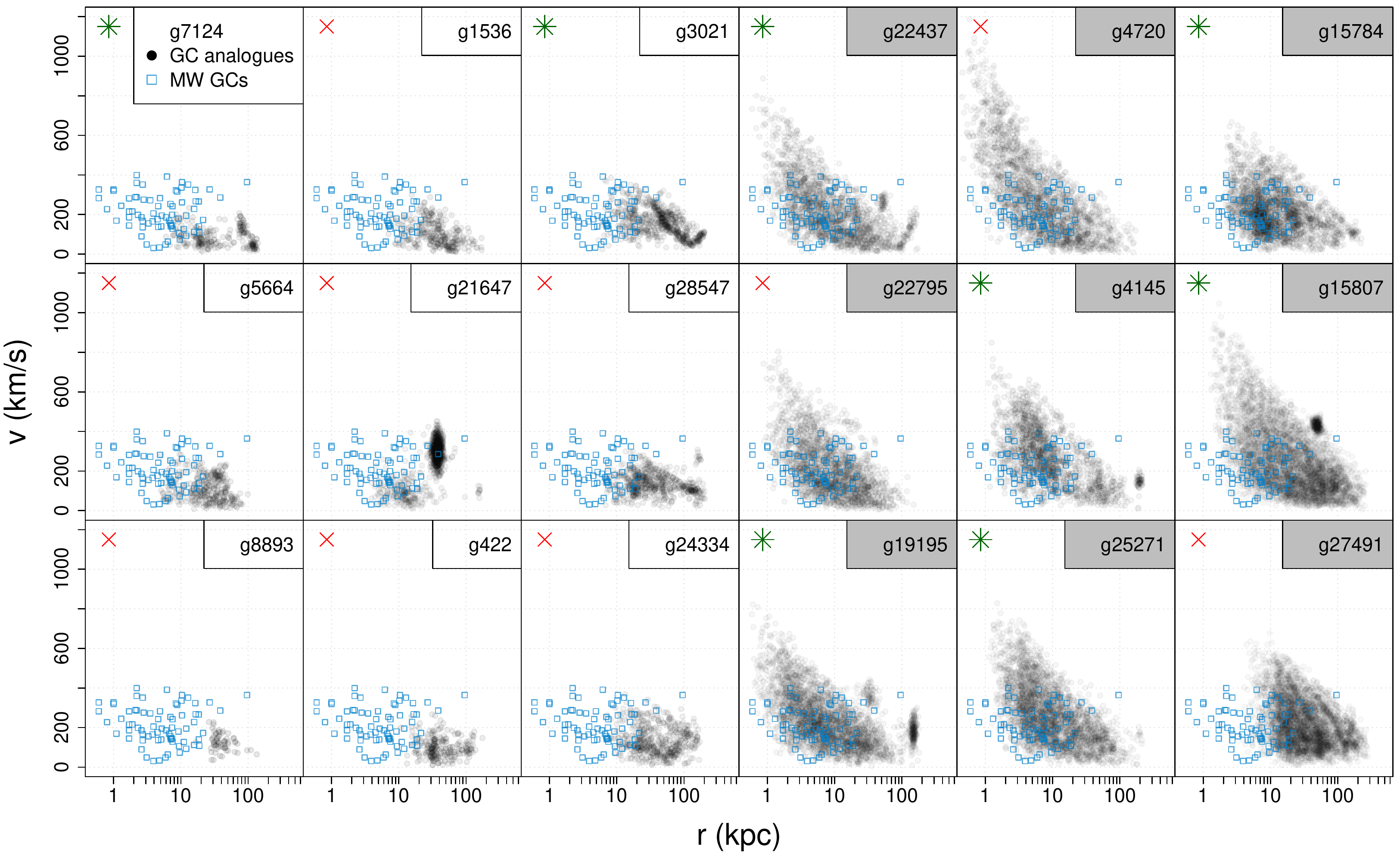}
	
	\caption{Velocity profiles of GC analogues from MUGS2 simulated galaxies, with MW GC velocities (for GCs with complete data) overplotted in hollow blue squares. Legends with a grey background correspond to the unregulated galaxies. Galaxies labelled with a green star are those whose total mass was estimated within the 95\% c.r., and those with a red ``X'' were not estimated as well (Section~\ref{sec:results4}).}
	\label{fig:velprofiles}
	
\end{figure*}

	The total velocities of the GC analogues as a function of their galactocentic distance are shown in Figure~\ref{fig:velprofiles}. For comparison, real MW GCs for which we have complete velocity information are overplotted as solid blue squares. Because much of the MW GC data are incomplete, the GCs shown represent roughly half of the total GC population in the MW.

	Three observations are immediately apparent in Figure~\ref{fig:velprofiles}: (1) the GC analogue velocity profiles of the MUGS2 galaxies sometimes differ substantially from that of the MW GC's, (2) some velocity profiles have unique clustered features that may represent satellites or recent mergers, and (3) there are often many more GC analogues than GCs in the MW. The number of GC analogues per MUGS2 galaxy ranges from 64 to 5106 (Table~\ref{tab:numberoftracers}), and is a reflection of the different star formation rates at high redshift. That is, the galaxies with higher numbers of GC analogues had more star formation in earlier times.

	Because the GC analogue population sizes differ from the MW GC population, which consists of 157 known GCs \citep{1996harrisPaper, 2010Harris}, we randomly sampled GC analogues to obtain the same sample size as the MW. Although it is possible that the galaxies with a larger number of GC analogues are larger galaxies, we did not use this as prior information in our analysis.

	\subsection{Creating mock heliocentric observations}\label{sec:mockobs}

	We created mock heliocentric observations of the GC analogues, such as would be viewed from a Sun-centered reference frame. This involved a series of steps, including transforming positions and velocities into a heliocentric frame, and introducing missing data and measurement errors to simulate real observations.

	\subsubsection{Transforming from galactocentric to heliocentric}
	
	The kinematic information of the GC analogues is in a galactocentric, Cartesian coordinate system with positions $(x,y,z)$ and velocities $(v_x, v_y, v_z)$ (i.e. $U$, $V$, $W$). We first transform the galactocentric positions into galactic coordinate $l$ and $b$ and heliocentric coordinates of right ascension  and declination.
		
	To create heliocentric velocities, we perform the inverse of the transformation provided in \cite{johnson1987}, adjusting for the solar motion \citep[we use the value from][]{schonrich2010}. The use of the same solar motion for every simulated galaxy will not affect the results because this quantity is treated as fixed and known in our analysis. That is, the same value for the solar motion is used when the observations are transformed back into the galactocentric frame.
		
	The above steps were performed on the MUGS2 GC analogue data to create perfect (i.e. without error) heliocentric mock data. To check the transformation, we converted these values back to the galactocentric frame using the relevant code in \textsc{GME}. There are discrepancies of $\simeq 2\times 10^{-13}$\kms ~in the velocities at small $r_{gc}$ when the transformation is performed \emph{without} scatter, due to floating point roundoff. However, the symmetry of these discrepancies and their tiny values will not contribute to any systematic bias in the result.

	\subsubsection{Introducing Measurement Uncertainties}\label{sec:measureuncert}

	Referring back to Eq.~\ref{eq:normal}, we call any difference between a measured value $x$ and the true value $\mu$ the \emph{error}. In order to analyse the MUGS2 data in a way that is most similar to the MW analysis, we must create realistic, observational errors. We achieve this by setting the mean $\mu$ to the true value of the quantity (e.g. $r$, $v_{los}$, etc) in the MUGS2 data, deciding on a value for $\sigma^2$, and drawing from the normal distribution determined by these parameters. How we choose to define $\sigma^2$ for each quantity ($r, v_{los}, \mu_{\alpha}\cos{\delta},\mu_{\delta}$) determines how much leverage a data point has on the final analysis.
	
	The galactocentric distances $r$ were assigned a measurement uncertainty of 5\% \citep[see][]{1996harrisPaper, 2010Harris}. The proper motion and line-of-sight velocity measurement uncertainties were drawn with replacement from the real data uncertainties by randomly selecting a row from the MW GC list given in Paper 2. 
	The proper motions uncertainties range from 0.03 -- 1.8 $\text{mas}~\text{yr}^{-1}$, and the line-of-sight velocity uncertainties range from 0.1 -- 15 $\text{km}~\text{s}^{-1}$. We excluded two large measurement uncertainties in this process --- those of Pal 3 and NGC 6218 --- to avoid assigning very large observed proper motions to the GC analogues that we deemed to be unrealistic. (In the MW analysis, we treated the proper motion of Pal 3 as unknown because the measurement uncertainty was so large.) We did investigate various distance measures to match MW GCs to GC analogues, so that GC analogues could be assigned uncertainties that were similar to their MW counterparts, but found these procedures gave final error distributions that were indistinguishable from the simple random sampling.

	After performing reference frame transformations and introducing measurement errors for the GC analogues of each MUGS2 galaxy, we subsampled the mock data to mimic the sample size of the MW's GC population. We randomly select 157 GC analogues from each MUGS2 galaxy, except in the case of $g8893$ for which there are only 64 GC analogues.
	
As noted previously, the velocity profiles of the GC analogues as a function of galactocentric distance are not always similar to the MW's GC profile (Figure~\ref{fig:velprofiles}). This is reflected in the subsamples' number density as a function of distance as well.

		\begin{figure}
	\centering

	\gridline{\fig{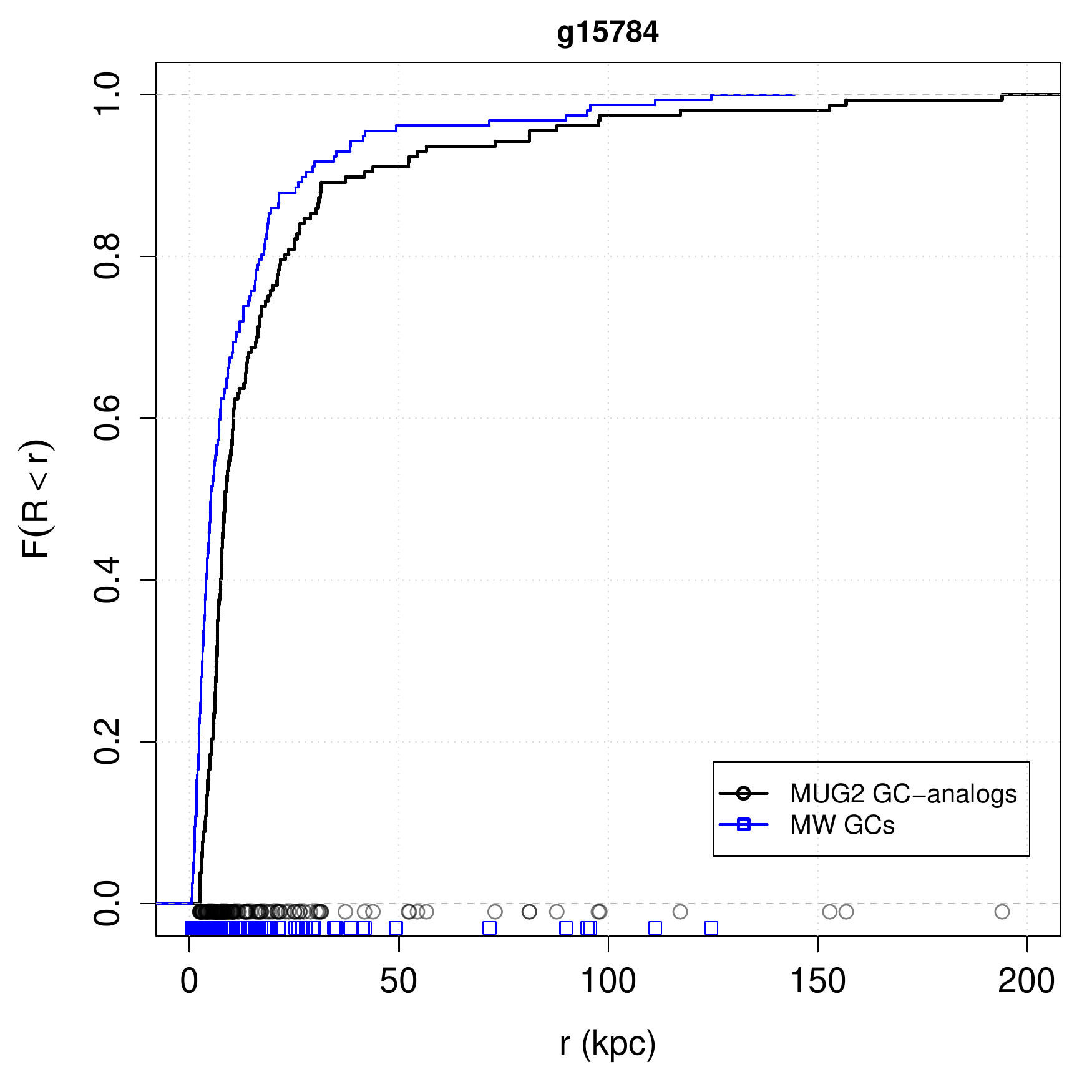}{0.47\textwidth}{(a) GC analogue empirical CDF for unregulated galaxy $g15784$.}
}

\gridline{
	\fig{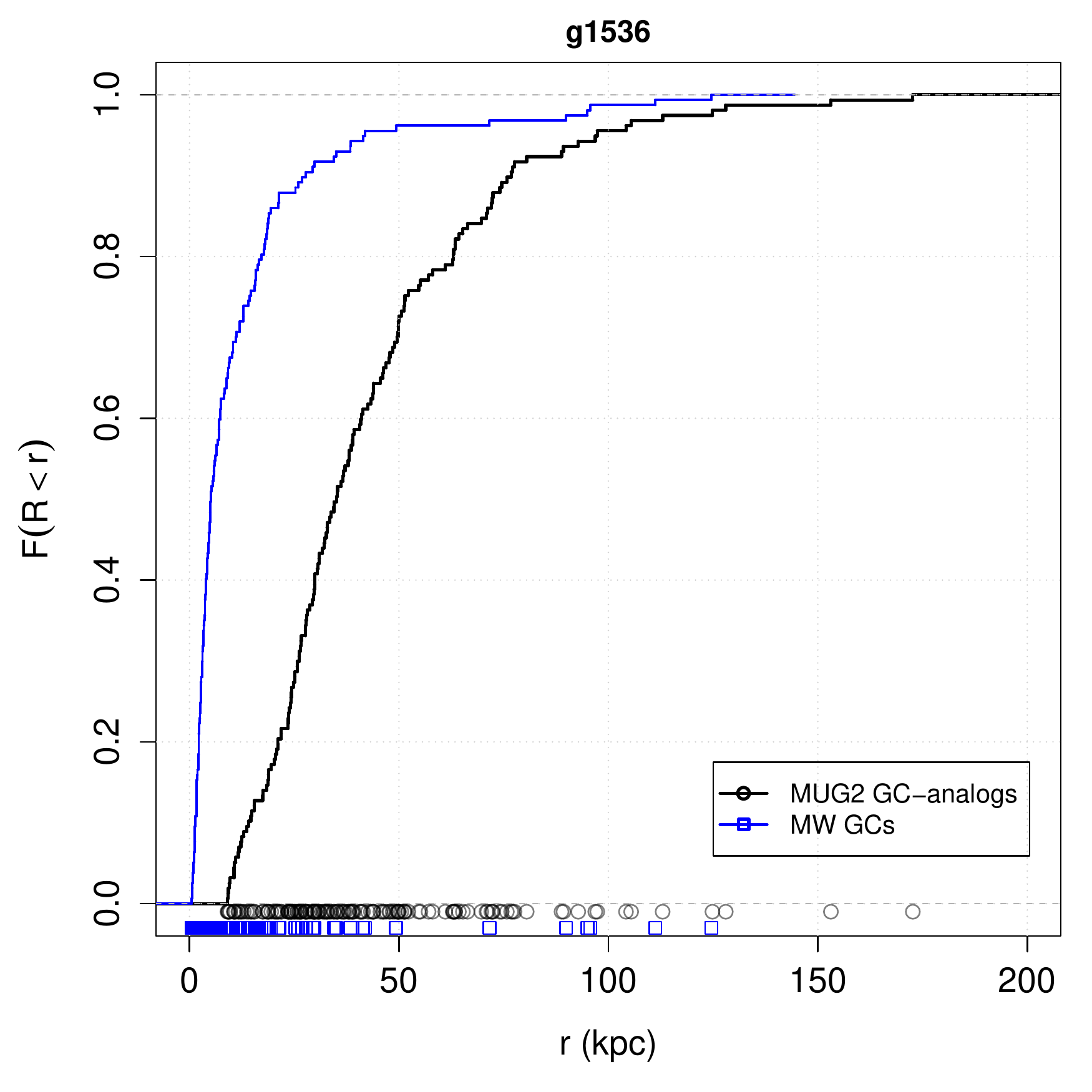}{0.47\textwidth}{(b) GC analogue empirical CDFs for regulated galaxy $g1536$.}
}
	
\caption{The black lines show the empirical CDFs for the subsampled GC analogue data from MUGS2 galaxies (a) $g15784$, an unregulated galaxy, and (b) $g1536$, a regulated galaxy. The blue lines show the empirial CDF for the MW GC data. The points along the bottom show the positions of individual GC analogues (top) and MW GCs (bottom). The empirical CDFs of other unregulated and regulated galaxies' are similar to these counterparts.}
	\label{fig:numdensity}
\end{figure}

Figure~\ref{fig:numdensity} shows the empirical cumulative distribution functions (CDFs) of the galactocentric distances $r$ of the subsampled GC analogue mock data for (a) $g15784$ and (b) $g1536$, as they compare to the empirical CDF for the MW GC system. The CDFs are calculated using the function \emph{ecdf} in the \emph{stats} package of the \textbf{R} Statistical Software Environment \citep{R}. The black curves are the MUGS2 GC analogues, and the blue curves are the MW GCs. The points along the bottom are the distances of the individual GCs from the Galactic center (black circles for MUGS2, blue squares for MW GCs). The empirical CDFs of $g15784$ and $g1536$  are quite different, with the former being much more similar to the MW's GC population CDF, especially at smaller $r$ (Figure~\ref{fig:numdensity}). The same observation is made for the other unregulated galaxies too; their CDFs appear more similar to the MW.

	\subsubsection{Creating Incomplete Data}\label{sec:createincomplete}
	
	Of the 157 GCs in the MW listed in Paper 2, 85 do not have proper motion measurements and 14 of this subset also lack line-of-sight velocity measurements. Within 20kpc of the MW centre, approximately 50\% (67/135) of the data are missing proper motion measurements, and beyond this distance approximately 87\% are missing proper motions. We mimic this distribution of incomplete data in the subsamples by removing 50\% of the proper motions within $20$kpc, and removing 87\% outside of this distance.
	
	In many of the MUGS2 regulated galaxies, any given subsample of 157 GC analogues of the simulated data resulted in very few points within 20kpc of the galactic centre. For example, drawing 157 samples from $g1536$ resulted in $\sim30$ GC analogues residing within 20kpc. This is in stark contrast to the MW, which has 135 GCs within $20$kpc. In the MW, proper motion measurements are available for at least 67 GCs within 20kpc. Removing 50\% of the proper motion measurements within 20kpc for $g1536$ therefore seems unrealistic. Thus, we decide to keep all but two proper motion measurements within 20kpc for $g1536$. All proper motions beyond $50$kpc are removed because we have only one complete data point past this distance in the MW. This procedure was needed for most regulated galaxies because they have few GC analogues within 20kpc.
	
	In Papers 2 and 3, we used the 14 GCs that lacked line-of-sight measurements to define the prior distribution in the number density profile parameter $\alpha$. For the MUGS2 GC analogues, we randomly remove 14 line-of-sight velocities and use the positions of these objects in the same way as we did for the real data.

	\section{Results \& Discussion}\label{sec:results4}
	
	We now apply GME to the GC analogue subsamples from each MUGS2 galaxy, and use the posterior distribution of model parameters to estimate $r_{200}$ and $M_{200}$ of each galaxy. The radius $r_{200}$ is defined as the distance from the galactic centre within which the mean mass density is 200 times the critical density of the universe. We use a Hubble constant of $73\kms\text{Mpc}^{-1}$, the same value used by \cite{keller2016} to create the galaxies\footnote{We also tried using the  \cite{Planck2015} result $H_0=67.8\kms\text{Mpc}^{-1}$ without significant changes to the results.}.
	
	We begin with a detailed look at $g15784$ and $g1536$ (Section~\ref{sec:testcases}) before showing the summarized results from all MUGS2 galaxies (Section~\ref{sec:MWclones}).

	\subsection{Detailed Cases: $g15784$ and $g1536$}\label{sec:testcases}
	
	Galaxies $g15784$ and $g1536$ were chosen for the initial analysis because they have quiet merger histories and represent typical examples of the regulated and unregulated populations of galaxies from the MUGS2 simulations. Additionally, $g15784$ lacks any strange features in its velocity profile (Fig.~\ref{fig:velprofiles}), has many GC analogues from which to draw samples (Table~\ref{tab:numberoftracers}), and its mock image of the galaxy appears MW-like (see Fig.~\ref{fig:galimages}). The regulated galaxy $g1536$, which is less concentrated than $g15784$, then makes for an interesting comparison.
	 
	\subsubsection{Parameter Estimates and Total Mass}

	The mean estimates of the model parameters given by the posterior distribution for galaxy $g15784$ are presented in Table~\ref{tab:prelimresults}, where the numbers in brackets represent the bounds of the 95\% marginal c.r. 
	
	The constant anisotropy parameter $\beta$ is more accurately estimated for $g15784$ than $g1536$, with the true values being approximately 0.6 and 0.8 respectively (see also Section~\ref{sec:parameters} and Figure~\ref{fig:betaestimates}).

	\begin{table}[b]
		\centering
		\caption{Model Parameter Estimates and Derived Quantities: $g15784$}
		\label{tab:prelimresults}
		\begin{tabular}{ccc}
			\hline
			\hline
			\T
			Parameter & Mean & 95\% Marginal c.r. \\
			\hline
			\T
			$\Phi_o (10^4\text{km}^{2}\text{s}^{-2})$ & 47 & (40, 57)  \\
			$\gamma$ & 0.41 & (0.30, 0.57) \\
			$\alpha$ & 3.04 & (3.02, 3.06) \\
			$\beta$ & 0.54 & (0.41, 0.66) \\
			\hline
			\hline
			\T
			Derived Quantity & & \\
			\hline
			\T
			$r_{200}$ (kpc) &  203 & (173, 232) \\
			$M_{200}$ ($10^{12}\msun$) & 1.1 & (0.6, 1.6) \\
			\hline

		\end{tabular}
	\end{table}

	The mean $r_{200}$ and $M_{200}$ of galaxy $g15784$ as predicted from the hierarchical Bayesian analysis are $r_{200}=203~(173, 232)$~kpc and $M_{200} = 1.1~(0.6, 1.6)\times10^{12}\msun$, where numbers in brackets are 95\% Bayesian c.r. (Table~\ref{tab:prelimresults}). These values are strikingly accurate--- the true values from the simulations are $220\kpc$ and $1.3\times10^{12}\msun$. 
	
	The results for $g1536$ are shown in Table~\ref{tab:prelimresultsg1536}. In this case, the method did not perform as well. The true $r_{200}$ and mass are 174kpc and
	$0.65\times 10^{12} \msun$, whereas the predicted values were $152~ (136, 170)$~kpc and $ 0.4~ (0.3, 0.6) \times 10^{12} \msun$.

	\begin{table}
		\centering
		\caption{Model Parameter Estimates and Derived Quantities: $g1536$}
		\label{tab:prelimresultsg1536}
		\begin{tabular}{ccc}
			\hline
			\hline
			\T
			Parameter & Mean & 95\% Marginal Credible Region \\
			\hline
			$\Phi_o (10^4\text{km}^{2}\text{s}^{-2})$ & 24 & (17, 35) \\
			$\gamma$ & 0.40 & (0.30, 0.61) \\
			$\alpha$ & 3.02 & (3.01, 3.04) \\
			$\beta$ & 0.51 & (0.28, 0.70) \\
			\hline
			\hline
			\T
			Derived Quantity & & \\
			\hline
			\T
			$r_{vir}$ (kpc) & 152 & (136,170) \\
			$M_{200}$ ($10^{12}\msun$) & 0.4 & (0.3, 0.6) \\
			\hline
			
		\end{tabular} 
		
	\end{table}

	\subsubsection{Cumulative mass profiles}
\begin{figure}
		\centering
		\gridline{
			\fig{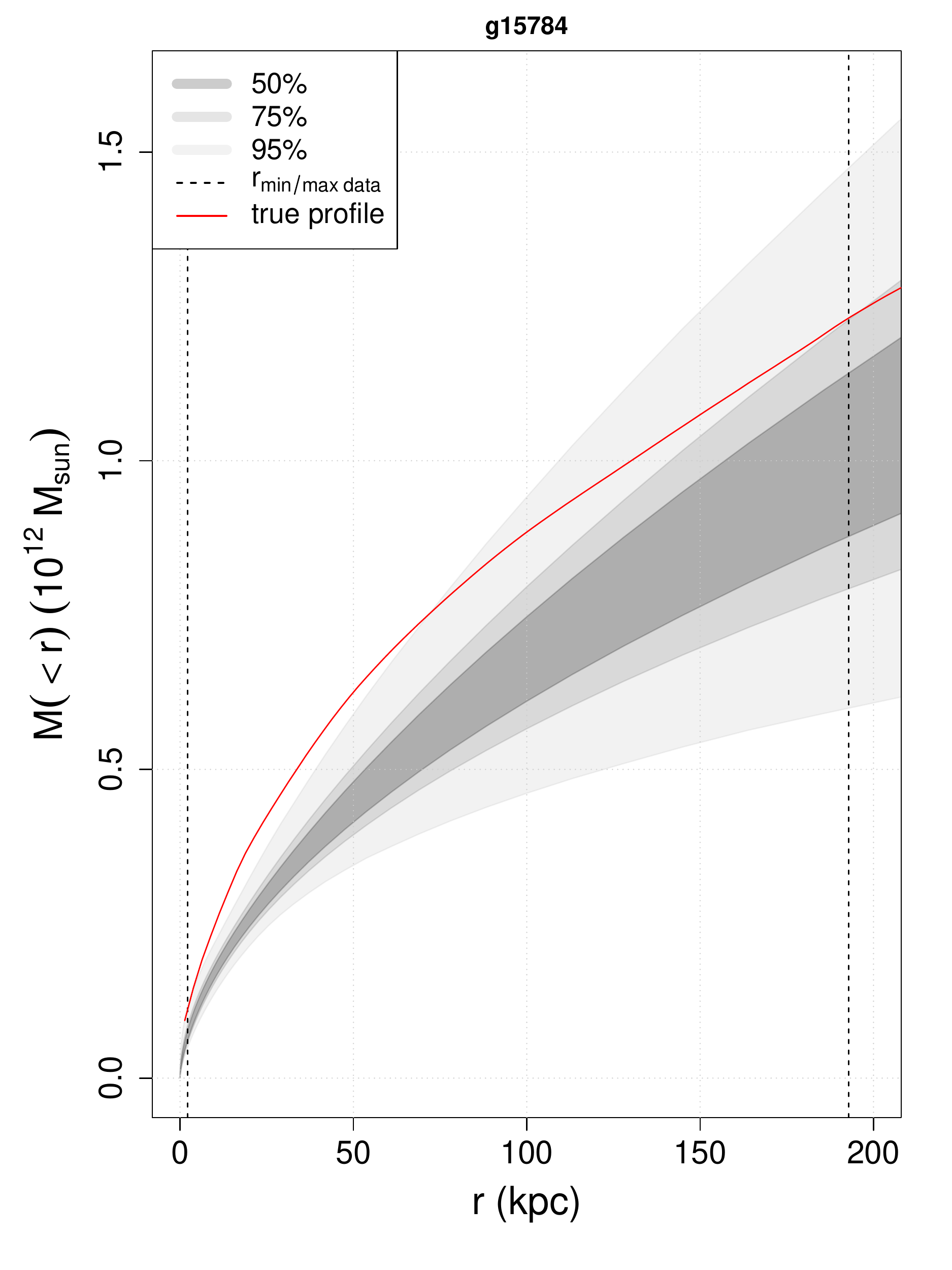}{0.4\textwidth}{(a) Cumulative mass profile for $g15784$}
		}
		\gridline{
			\fig{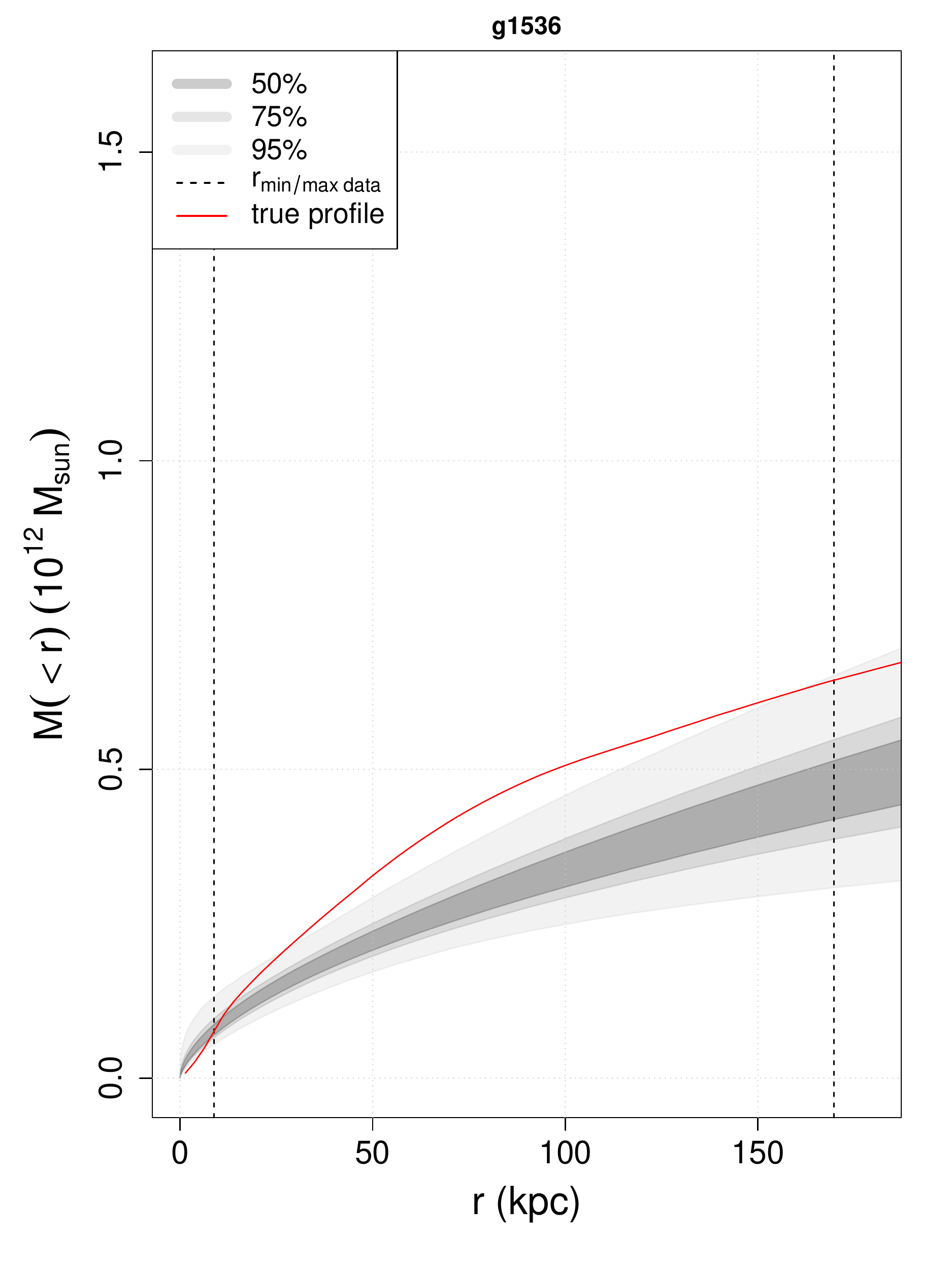}{0.4\textwidth}{(b) Cumulative mass profile for $g1536$}
		}
		\caption{The predicted (grey-shaded regions) and true (red curve) cumulative mass profiles for galaxy (a) $g15784$ and (b) $g1536$. The dashed-vertical lines indicate the range of the mock observations.} 		\label{fig:prelimresults}
		
	\end{figure}

	The cumulative mass profiles with Bayesian c.r.~for each galaxy, using Equation~\ref{eq:Mr4} and the samples from the posterior distributions, are shown in Figure~\ref{fig:prelimresults}. The grey-shaded regions indicate the 50, 75, and 95\% c.r., and the true cumulative mass profiles are the solid red curves. The vertical-dashed lines show the range of the subsampled data used in the analysis.
	
	The predicted $M(<r)$ profile for both $g15784$ and $g1536$ falls below the true profile for most values of $r$. That is, even though the total mass is well estimated for $g15784$, the predicted mass profile shows disagreement with the true mass profile in Figure~\ref{fig:prelimresults}(a). In both cases, the model fit appears to be a compromise between the inner and outer regions of the galaxy. One notable difference between the two galaxies is that the mass within 10kpc is underpredicted for $g15784$, but overpredicted for $g1536$. Additionally, the true cumulative mass profile for $g1536$ has a different shape than the predicted one.

	\subsubsection{Specific energy profiles}\label{sec:energyprofiles}
	
	The hierarchical Bayesian method treats the true positions and velocities as nuisance parameters, sampling them in the MCMC hybrid-Gibbs algorithm (see Papers 1,2, and 3). As a result, we obtain marginal posterior distributions for the galactocentric velocity and position of each GC analogue.
	
	In Paper 3,  we used these distributions to estimate the specific energy $\mathcal{E}$ of each tracer, given the mean model parameters. We compared these estimated energies to energies calculated from the actual measurements of position and velocity and the model parameters. By looking at the energies as a function of galactocentric distance $r$, we noted that GME attempts to reconcile outlier GC energies in light of the other GCs' energies.
	
	Here, we have the luxury of comparing the estimated energies to the true energies of the GC analogues given the actual gravitational potential calculated from the MUGS2 simulations.
	
\begin{figure*}

		\gridline{
			\fig{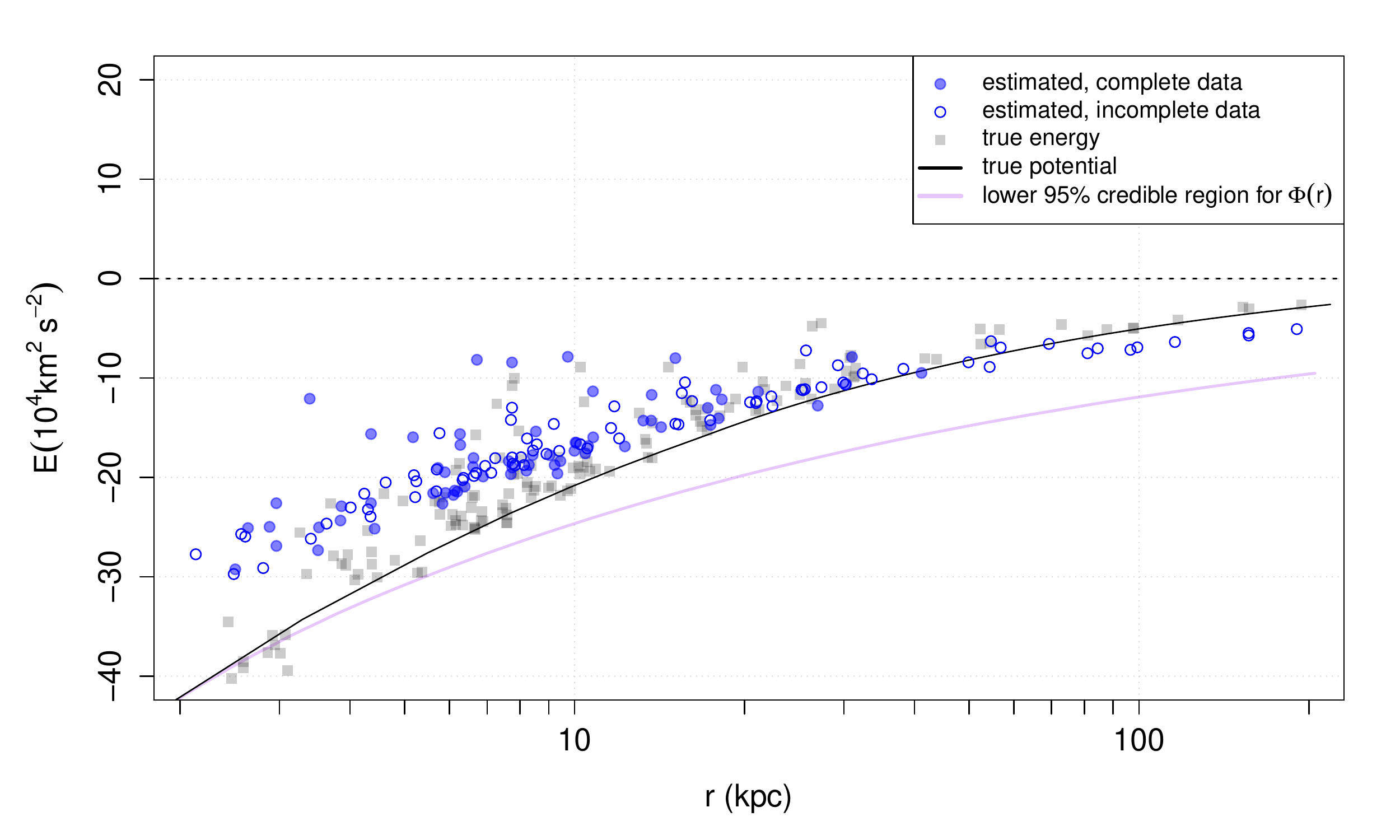}{0.8\textwidth}{(a) $g15784$}
		}
		\gridline{	
			\fig{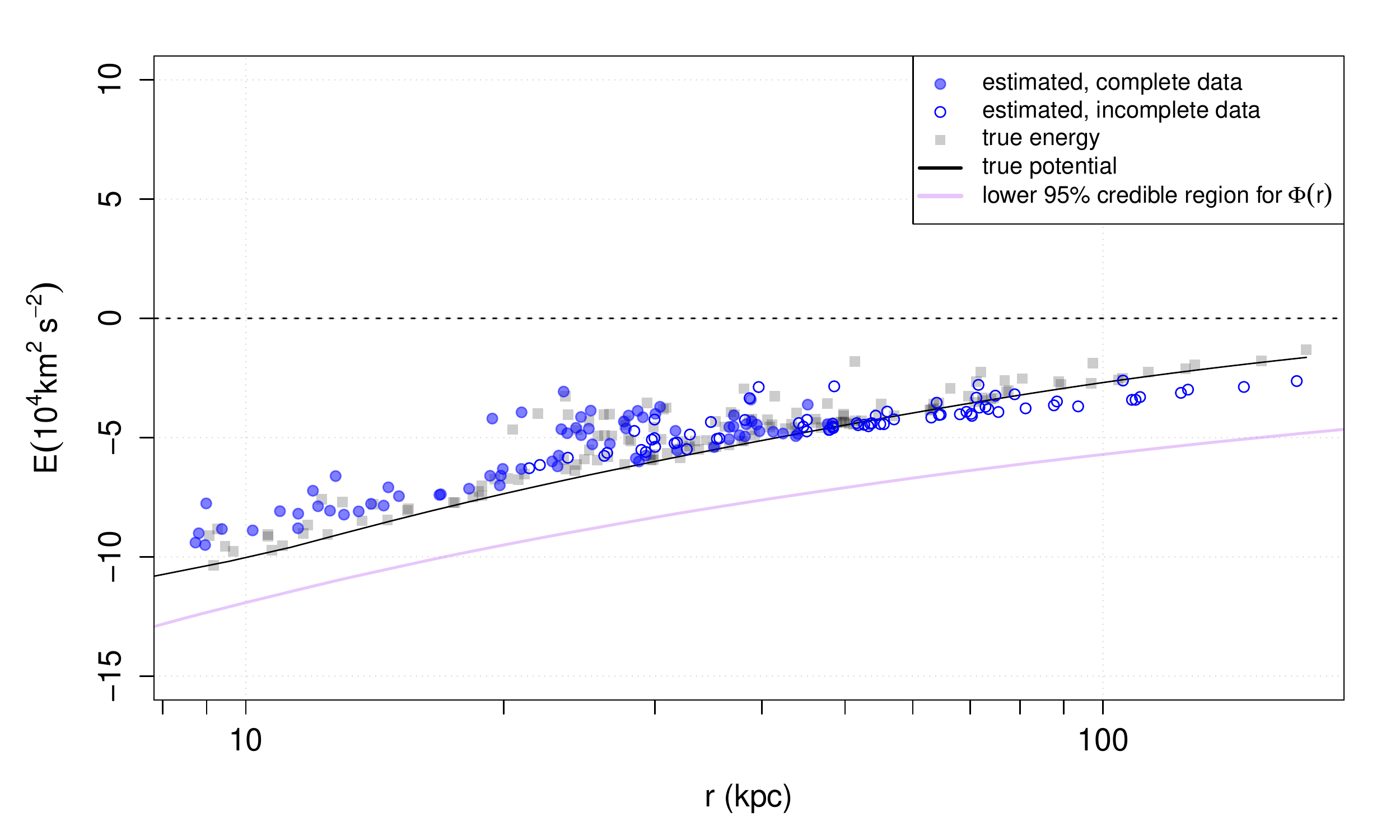}{0.8\textwidth}{(b) $g1536$}
		}
				\caption{Energy profile of GC analogues from (a) $g15784$ and (b) $g1536$. The solid blue and open blue circles are the estimated energies, calculated using the mean values of both the model and nuisance parameters of the posterior distribution. The grey squares show GC analogues' true energies, calculated using the true kinematic values and the true gravitational potential. The purple curve is the lower bound of the 95\% c.r. for the predicted gravitational potential, and the solid black line is the true gravitational potential.}
				\label{fig:eprofiles} 
	
\end{figure*}

	Figure~\ref{fig:eprofiles} shows the true specific energies (grey squares) and estimated specific energies (blue circles) of the GC analogues from $g15784$ and $g1536$.  The true energies are calculated using the position and velocity data direct from the simulations, and the true gravitational potential at their distances. The estimated energies are calculated from the mean values of the nuisance parameters ($r, v_{los}, \mu_{\alpha}\cos{\delta}, \mu_{\delta}$) provided by the posterior distribution samples, and the mean estimates of the model parameters. The estimates for the \emph{incomplete} data are open circles, while the estimates for the \emph{complete} data are filled.
	
   In both galaxies, the energy estimates appear to display statistical shrinkage; the free parameters for the velocity and positions allow the estimated $\mathcal{E}$ to move toward a common curve in $\mathcal{E}-r$ space. Overall, this ability of our method to adjust the energy values of the GC analogues reflects the result found in Paper 3, for the real MW data.

	Figure~\ref{fig:eprofiles} also reveals a disagreement between the estimated energy profile and the true energy profile of the tracers. The differences are largest at small and large $r$ for both galaxies. The differences are more extreme for $g15784$ than $g1536$, and this likely leads to the more uncertain mass estimate in the former (Figures~\ref{fig:prelimresults}). 
	
	The estimated energy profile for galaxy $g1536$ appears to match the true energy profile more closely than that of $g15784$, but this is not an indication of a good model fit. Disagreements at small and large $r$ still exist, and the cumulative mass profile in Figure~\ref{fig:prelimresults}(b) shows a poor and underestimated fit for $g1536$'s mass.

	Galaxy $g15784$ has incomplete data at all $r$, whereas almost all of the data within 20kpc of $g1536$ are complete. A high percentage of the data for $g1536$ are also complete between 20kpc and 50kpc (Figure~\ref{fig:eprofiles}(b)). Consequently, the inner GC analogues of $g1536$ will have the most leverage in the model fit.

 In Papers 2 and 3, we justified using a single power-law for the gravitational potential. Our argument was that at large distances a value of $\gamma=0.5$ provides a good approximation to an NFW \citep{nfw1996} gravitational potential. In the case of $g1536$, the assumed gravitational potential does not hold at all radii, and so an abundance of complete data in the inner regions of the galaxy has probably biased the mass estimate. Indeed the complete data within 20kpc correspond to the region in the mass profile that has the best mass prediction (Figure~\ref{fig:prelimresults}(b)).
 
 In the next section, we review the results for the rest of the MUGS2 galaxies, and detect a similar bias in the mass estimates of the other regulated galaxies.
		
		\vspace{10ex}
		
\subsection{Analysis of all MUGS2 galaxies}\label{sec:MWclones}
	
	\subsubsection{Parameter estimates}\label{sec:parameters}
	
\begin{figure*}
		\centering
		\includegraphics[scale=0.5, trim=0.7cm 0cm 0cm 0cm]{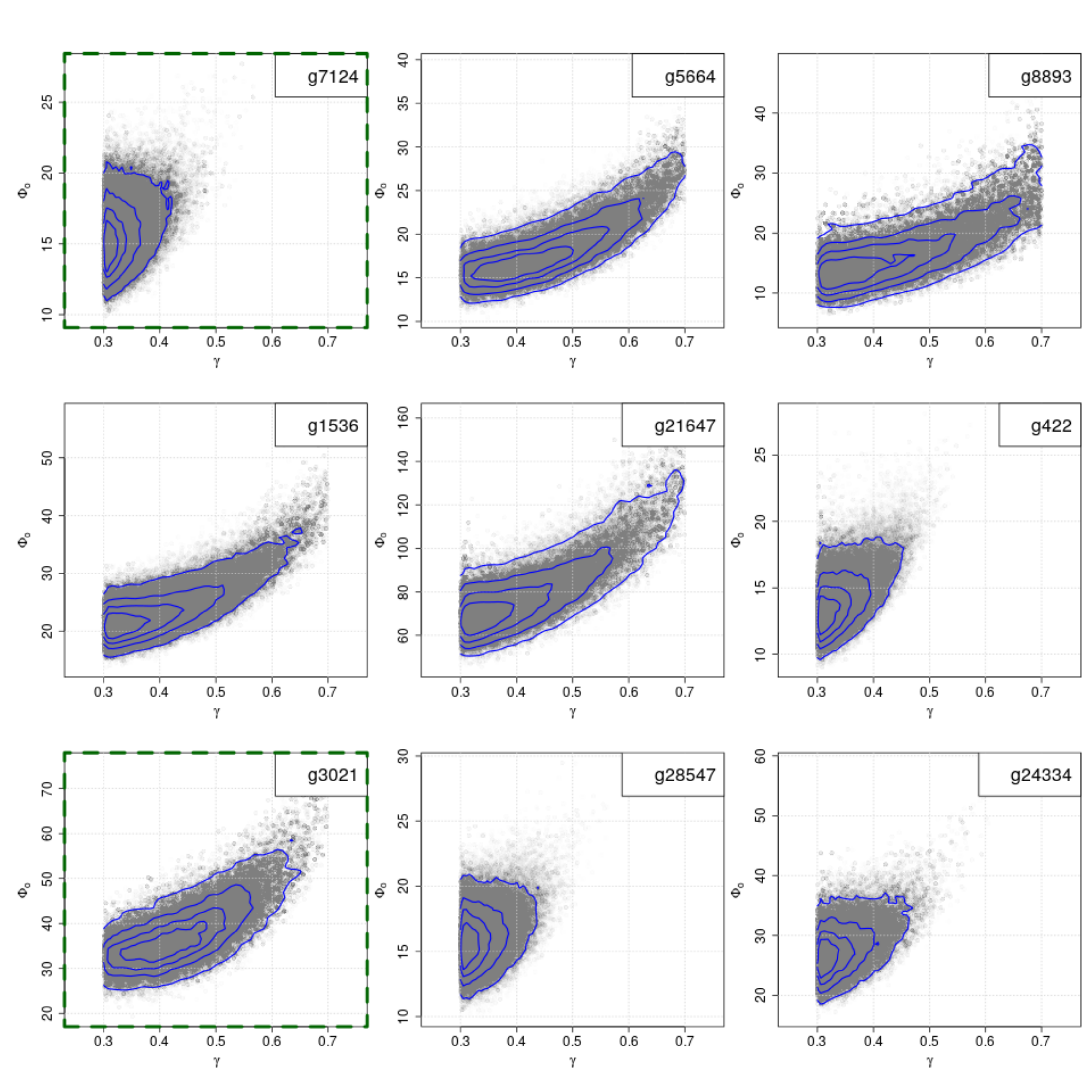}
		\caption{Joint distributions for $\Phi_o$ and $\gamma$, for the \emph{regulated} galaxies. Points are samples from the Markov chain and contours represent (from inner to outer) the 25\%, 50\%, 75\%, and 95\% Bayesian c.r.. The green-dashed boxes around the figure indicate that $M_{200}$ was correctly predicted within the 95\% c.r..}
		\label{fig:jointphigam_reg}
	\end{figure*}

\begin{figure*}		
		\centering
		\includegraphics[scale=0.5, trim=0.7cm 0cm 0cm 0cm]{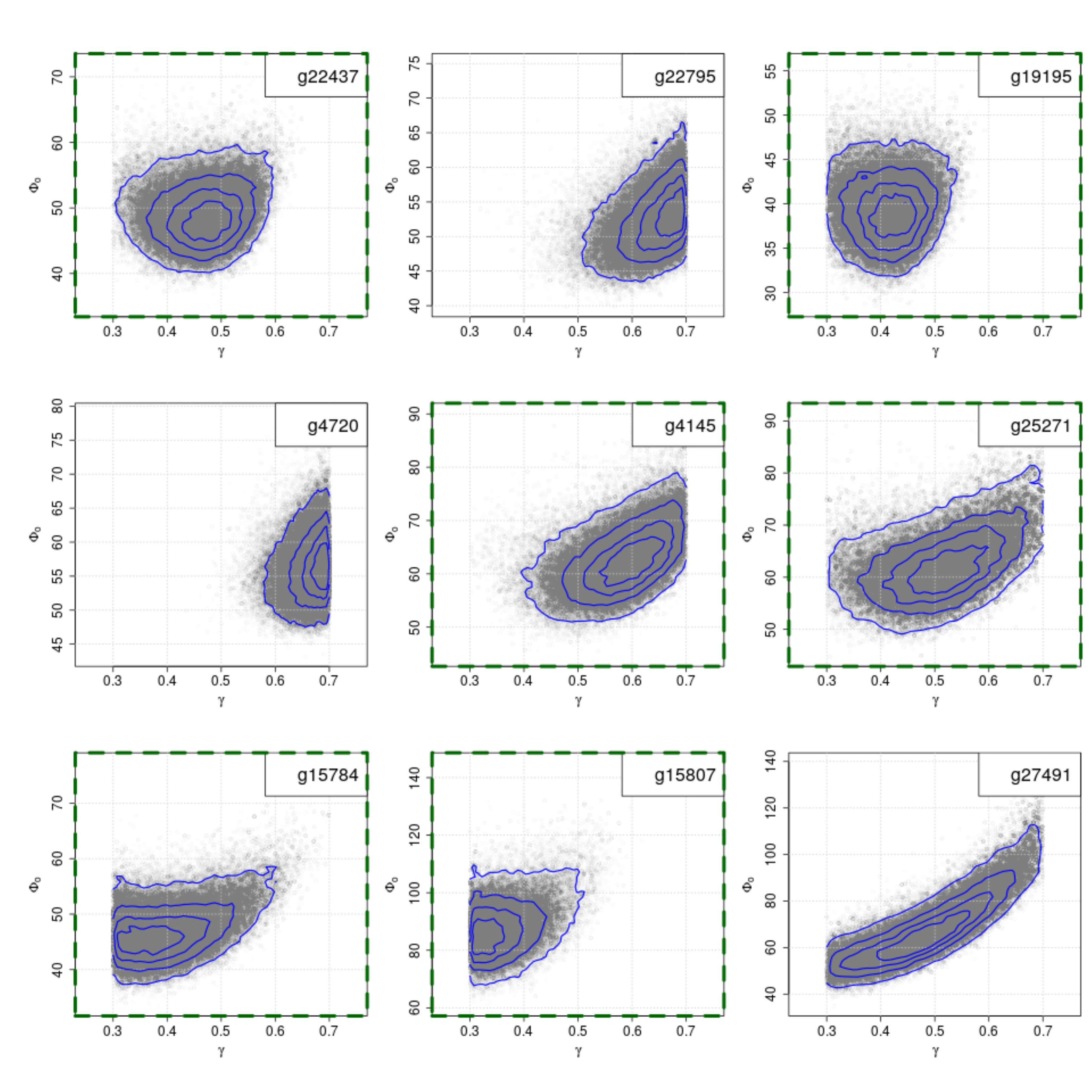}
		\caption{Joint distributions for $\Phi_o$ and $\gamma$, for the \emph{unregulated} galaxies. Points are samples from the Markov chain and contours represent (from inner to outer) the 25\%, 50\%, 75\%, and 95\% Bayesian c.r.. The green-dashed boxes around the figure indicate that $M_{200}$ was correctly predicted within the 95\% c.r..}
		\label{fig:jointphigam_unreg}
	\end{figure*}

	The joint posterior distributions for $(\Phi_o, \gamma)$ --- the model parameters of the gravitational potential --- are shown in Figures~\ref{fig:jointphigam_reg} (regulated galaxies) and \ref{fig:jointphigam_unreg} (unregulated galaxies). Green dashed boxes indicate galaxies whose total mass was estimated within the 95\% Bayesian c.r. (Section~\ref{sec:massresults}).
	
	For many regulated galaxies, the free parameter $\gamma$ attempts to reach a location in parameter space outside of the prior distribution. Such behaviour indicates the model struggled to accurately describe the gravitational potential given the data and prior assumptions. 
	
	The behaviour of $\gamma$ in the joint posterior distributions for the unregulated galaxies is also inconsistent. For example, the unregulated galaxies that are most underestimated and overconfident are $g22795$, $g4720$, and $g27491$. The joint distributions $p(\Phi_o, \gamma)$ in the first two cases show $\gamma \rightarrow 0.7$ (Figure~\ref{fig:jointphigam_unreg}), whereas for the distribution the latter is quite diffuse. 
	
	One interpretation of the results for $g22795$ and $g4720$ is that $\gamma$ is attempting to reach a value larger than 0.7 because of the massive bulges of these unregulated galaxies; in the limit that $\gamma \rightarrow 1$, the gravitational potential model goes to a Keplerian potential (Equation~\ref{eq:potential4}). In Paper 2 when we analysed the MW data, the parameters $\Phi_o$ and $\gamma$ appeared to be anticorrelated. Thus, a larger value of the latter leads to a smaller value of the former --- which leads to a smaller mass estimate. However, in the present analysis with MUGS2 data, the parameters appear correlated.  Therefore, we should be cautious about extrapolating conclusions from these blind tests to those about the MW that were arrived at using real GC data.
	
	A posterior distribution that is truncated by a prior distribution typically indicates a poor model fit. The only joint posterior distributions that look well-behaved in Figure~\ref{fig:jointphigam_reg} and \ref{fig:jointphigam_unreg} are those of $g22437$, $g19195$, and $g25271$. Notably, the true $M_{200}$ values for these unregulated galaxies are well within the 95\% Bayesian c.r. (Figure~\ref{fig:mvirMWclones} and Section~\ref{sec:massresults}).

\begin{figure*}
	\centering
	\includegraphics[scale=0.5]{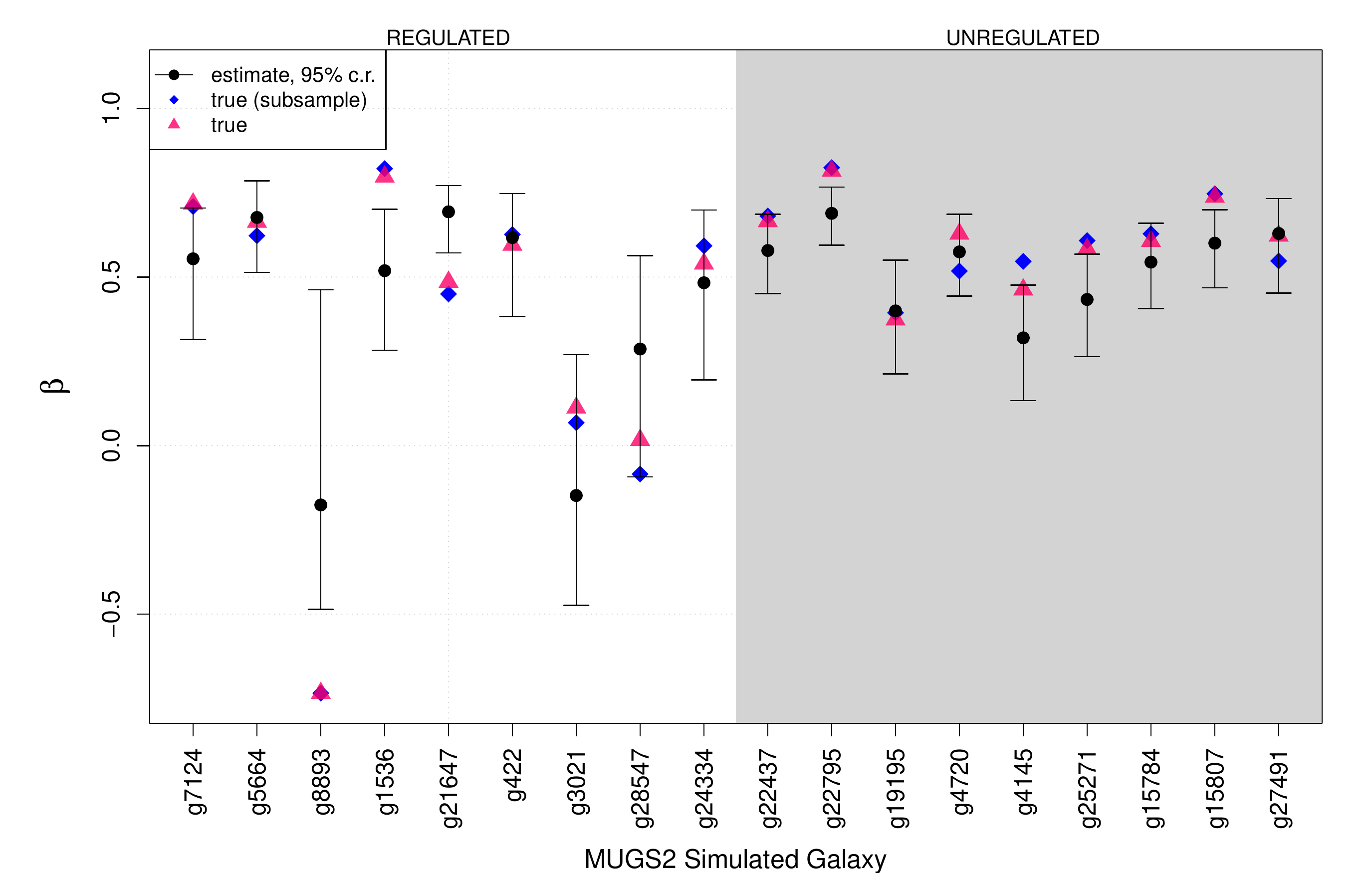}
	\caption{The median estimates (black points) of the constant anisotropy $\beta$, compared to the true constant anisotropy of the GC analogue subsample (blue diamonds) and the GC analogue population (pink triangles). The error bars represent the 95\% Bayesian c.r. from the marginal posterior distribution for $\beta$.}		\label{fig:betaestimates}
\end{figure*}

	We also obtain estimates for the constant anisotropy parameter $\beta$  of the tracer population. Figure~\ref{fig:betaestimates} shows the $\beta$ estimates (black circles) with 95\% c.r. (error bars), and the true constant anisotropy for (1) the GC analogue subsample (blue diamonds) and (2) the total GC analogue population (pink triangles). The unregulated galaxies are highlighted with a grey background.
		
	The true $\beta$ value for the GC analogue population is captured within the c.r.~11 out of 18 times. Although the mock data are incomplete beyond $50$kpc, making it difficult to constrain $\beta$, our estimates do not display significant bias when compared to the true constant anisotropies. 
		
			A feature of note is that galaxies $g4145$ and $g22795$ both have companions (Figure~\ref{fig:galimages}), and $\beta$ is very underestimated in both of these cases. Interestingly, galaxy $g19195$ has a noticeable  feature in $v-r$ space (Figure~\ref{fig:velprofiles}) but $\beta$ is accurately estimated. Also worthy of note is the poor estimate of $\beta$ for $g8893$, which not only has the smallest number of GC analogues (64) but is also one of the more irregularly shaped galaxies in Figure~\ref{fig:galimages}. Our prior distribution on $\beta$ did not allow for values less than -0.5, and yet the true $\beta$ fell outside this range.
			
		The regulated galaxies' $\beta$ estimates perform less well than those of the unregulated galaxies. In Section~\ref{sec:mergerhistory} and Figure~\ref{fig:mergerhistory}, we will see that the regulated galaxies have a higher percentage of incomplete tracer data, making it more difficult for the model to estimate the anisotropy of these tracer populations.	Nevertheless, the accuracy of the $\beta$ estimate does not appear to be related to the accuracy of $M_{200}$ or $r_{200}$, shown next.
		
	\subsubsection{Mass ($M_{200}$) and $r_{200}$ estimates}\label{sec:massresults}
	
\begin{figure*}[t]
	\centering
	\includegraphics[scale=0.6]{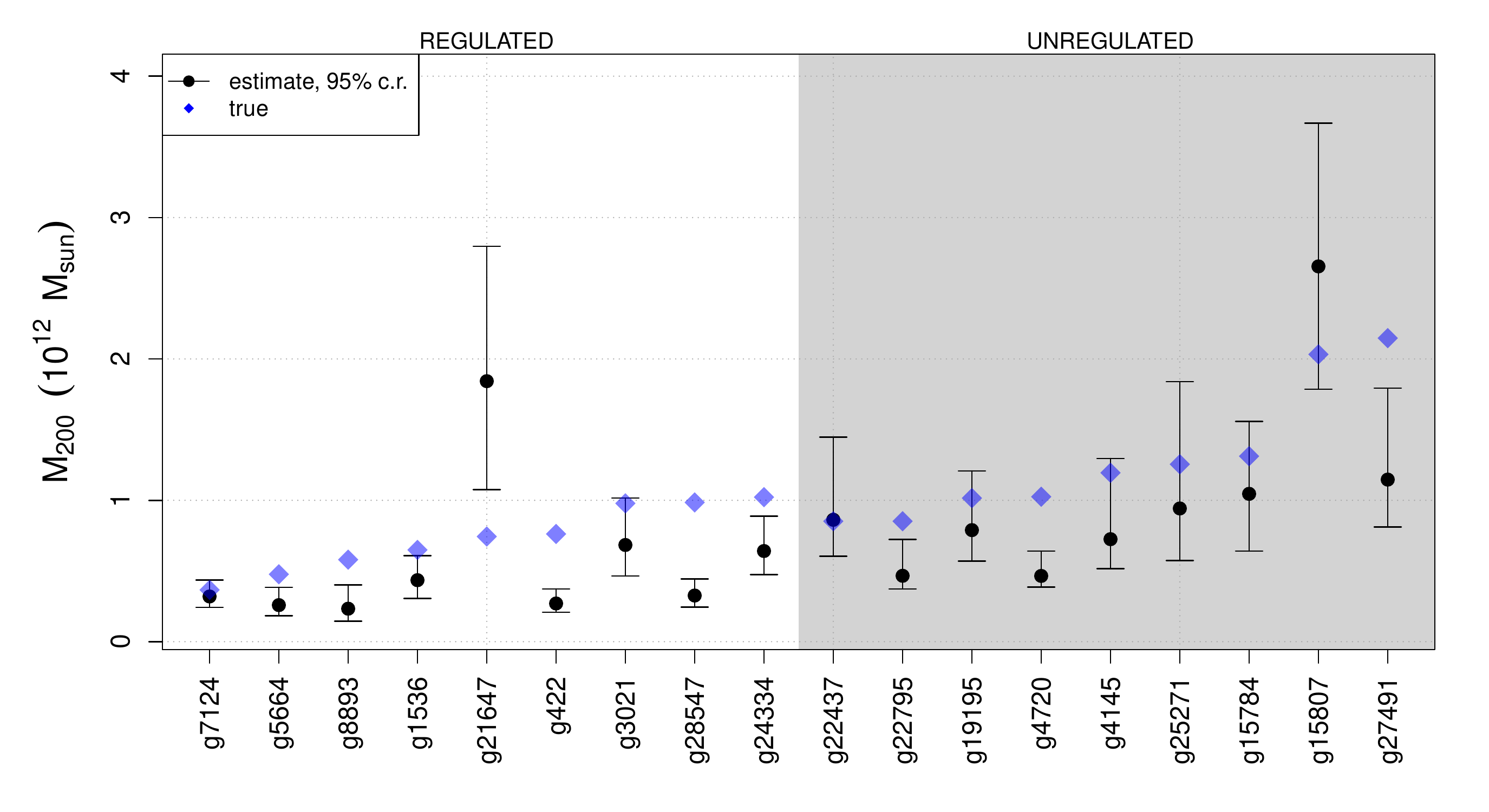}
	
	\caption{Total mass estimates of the MUGS2 galaxies when using a random subsample 157 of GC analogues. The estimates (black circles) and 95\% c.r. (erros bars) were calculated from the posterior distributions. The true $M_{200}$ values from the simulations (blue diamonds) show a tendency to be higher than the predictions.}
	\label{fig:mvirMWclones}
\end{figure*}

Figure~\ref{fig:mvirMWclones} summarizes the median estimates of the total mass ($M_{200}$) for all MUGS2 galaxies. The estimates are shown as black circles, the true values are blue diamonds, and the 95\% c.r.~are shown as error bars. The unregulated galaxies are shown with a grey background, and their $M_{200}$ estimates have notably wider marginal distributions than the other nine galaxies. Many of the regulated galaxies have total mass estimates that are both underestimated and overconfident.

\begin{figure*}[t]
	\centering
	\includegraphics[scale=0.6]{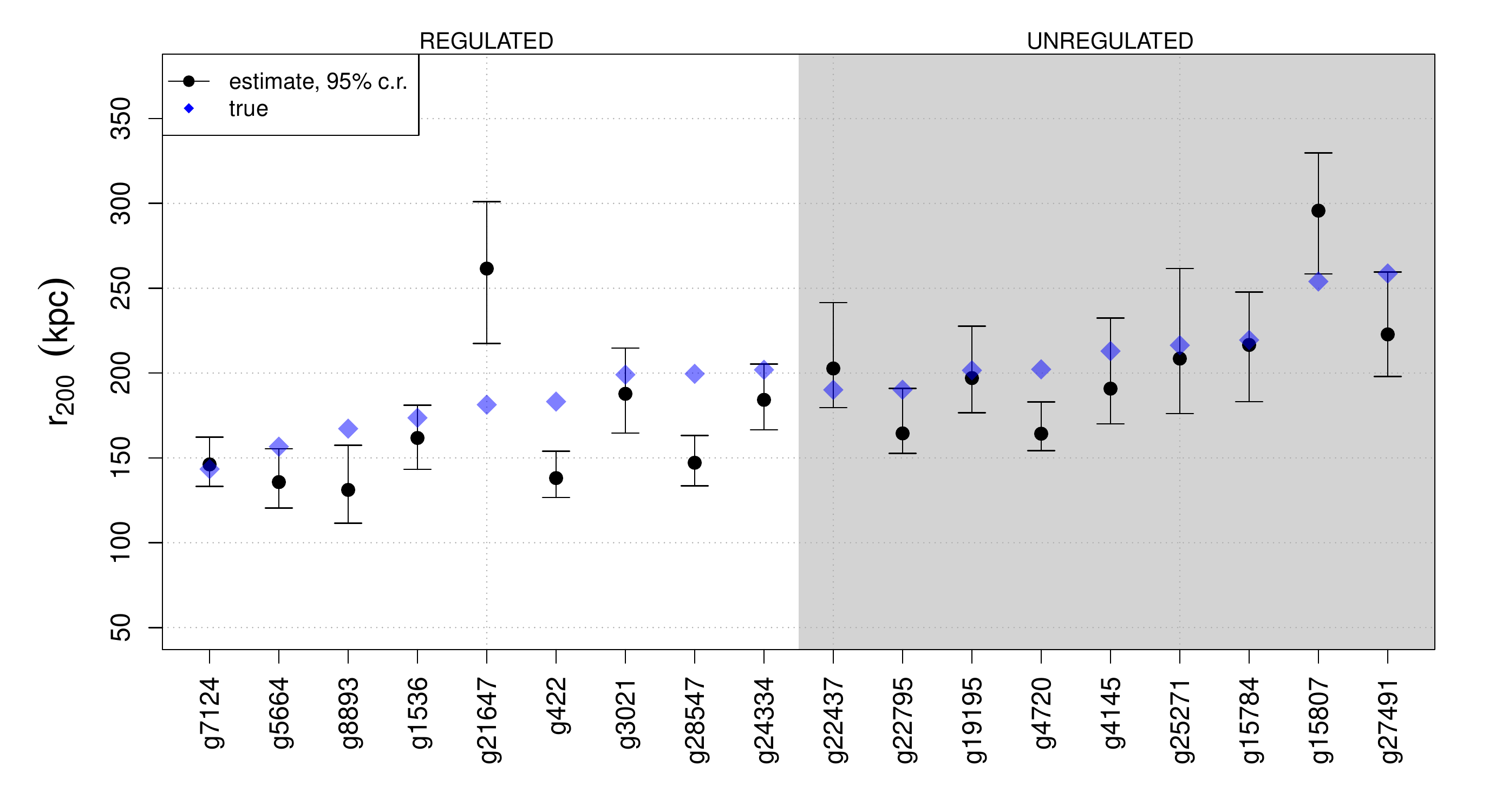}
	\caption{The median estimates of $r_{200}$ (black points), compared to the true values (blue diamonds). The error bars represent the 95\% Bayesian c.r..}
	\label{fig:rvirMWclones}
\end{figure*}

The true total mass is captured within the 95\% c.r. in 8/18 cases, with the unregulated galaxies having better coverage than the regulated ones. Overall, the $M_{200}$ values are underestimated by the median, with the exceptions of galaxies $g15807$ and $g21647$ (which are overestimated), and $g7124$ and $g22437$ (which are estimated quite well). The estimates and true values of $r_{200}$ are shown in Figure~\ref{fig:rvirMWclones}, and these echo the mass results.

In the next section, we investigate how the error in the mass estimates might be related to incomplete data, model assumptions, and the evolutionary history of the MUGS2 galaxies.

\subsubsection{Error in mass estimates}\label{sec:mergerhistory}

We now investigate why the method did not fair well in some cases, using the knowledge from the simulations. We know how many GC analogues are unbound from each galaxy, and from our mock observations we know the percentage of incomplete velocity data. Additionally,  we have information about each galaxy's merger history through the redshift of their last major merger ($z_{\text{lmm}}$) and the redshift at which they acquired half of their final mass ($z_h$) \citep{keller2016}.

\begin{figure*}
	\centering
	\includegraphics[scale=0.75, trim=0cm 1cm 0cm 0cm]{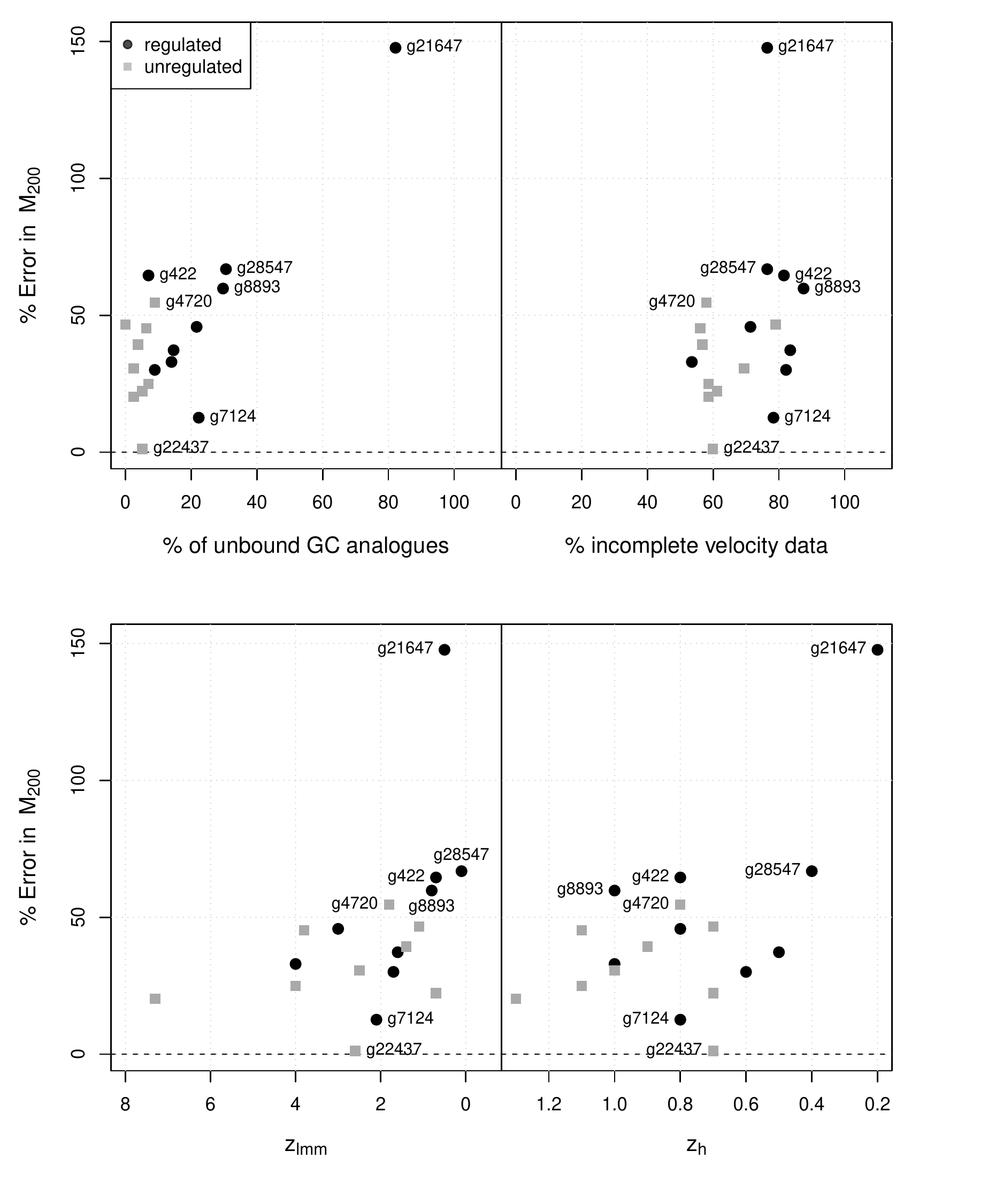}
	\caption{The absolute percentage error in $M_{200}$ versus the percentage of unbound particles, the percentage of incomplete data, the redshift of the last major merger ($z_{lmm}$), and the redshift when the galaxy had acquired half of its final mass ($z_h$). Regulated galaxies are solid black circles, and unregulated galaxies are grey squares.}
	\label{fig:mergerhistory}
	
\end{figure*}

Figure~\ref{fig:mergerhistory} shows the error in the mass estimates as a function of the following quantities: the percentage of unbound GC analogues, the percentage of incomplete data, $z_{lmm}$, and $z_h$. The absolute percent error in the total mass is calculated by
\begin{equation}
	\text{Percentage Error}  = |\frac{ \widehat{M_{200}} - M_{200} }{ M_{200} }| \times 100\%
\end{equation}
where $\widehat{M_{200}}$ is the estimate and $M_{200}$ is the truth.

The two galaxies with the highest absolute error ($g21647$ and $g28547$) are those that had the most recent major mergers and that have the highest percentage of unbound particles. Interestingly, these galaxies represent errors in two extremes --- one was severely overestimated, and the other underestimated. In general, however, there is only a slight trend for galaxies with recent major mergers to have the most inaccurate mass estimates (lower-left panel in Figure~\ref{fig:mergerhistory}).

On average, the regulated galaxies have a higher proportion of incomplete data and unbound GC analogues than the unregulated galaxies. This may explain why the velocity anisotropy parameter was more poorly estimated for the regulated galaxies than for the unregulated ones (Figure~\ref{fig:betaestimates}).

Overall, Figure~\ref{fig:mergerhistory} does not suggest any significant trends in the mass error with respect to merger history, percentage of incomplete data, or percentage of unbound particles.  Thus, except in extreme cases, these quantities may not play a role in underestimating the mass. In the following section, we explore the possibility that the inner GC analogues are influencing the model fit, as suggested in Section~\ref{sec:energyprofiles}.

\subsection{Sensitivity to Inner Tracers}\label{sec:sensitivity}
	
Our studies of the MW, and in particular the sensitivity analyses in Papers 2 and 3, found an increased mass estimate when inner GCs were not included in the analysis. We repeat this kind of analysis here by removing the inner GC analogues from the MUGS2 galaxies, and recalculating the mass estimate $M_{200}$.

Each galaxy in the MUGS2 simulation is unique in shape and spatial distribution of GC analogues. In order to remove inner GC analogues in a consistent way across all galaxies, we find the scale radius $r_s$ of each galactic disk via a simple exponential fit, and then remove GC analogues within $7.5r_s$. The MW's scale radius is roughly 2~kpc, so in our own Galaxy this cut is akin to removing GCs within $r_{\text{gc}}=15$~kpc of the Galactic center. It should be noted that the galaxies already had GC analogues removed within $3r_s$ in the original analysis.

\begin{figure*}
	\centering
	\includegraphics[scale=0.55, trim=5cm 0cm 5cm 0cm ]{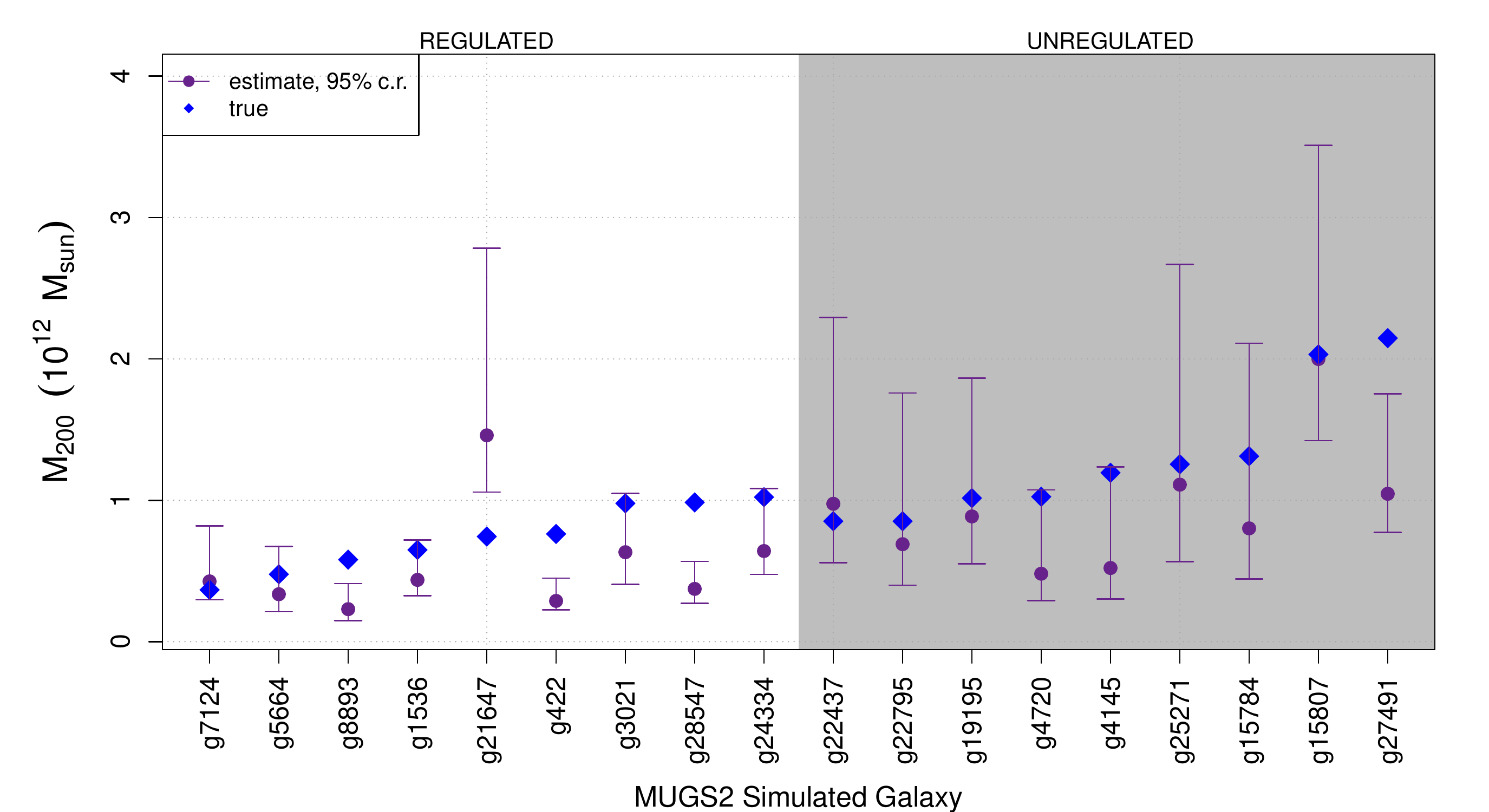}
	
	\caption{Results from the sensitivity test: mass estimates after selecting tracers outside of $r = 7.5r_s$). True values are shown as blue diamonds, estimates are dark purple circles, and the 95\% c.r. are shown as error bars.}
	\label{fig:sensitivityresults}
\end{figure*}

The new mass estimates after removing the inner GC analogues are shown in Figure~\ref{fig:sensitivityresults}. In general, the estimates are better but at the cost of wider Bayesian c.r. The true $M_{200}$ values lie within the Bayesian c.r. in 13 out of 18 cases--- a slight improvement from 8 out of 18 (Figure~\ref{fig:mvirMWclones}). Three regulated galaxy masses that were previously underestimated are now captured within the uncertainties. The mass estimate for $g15807$, previously a large outlier, is also notably improved.

In light of this result, and evidence presented in Section~\ref{sec:energyprofiles}, we suggest that the combination of the GC analogue number density profile, the percentage of complete data at small $r$, and the incomplete data at large $r$ are the culprits of the mismatch between the predicted and true mass profiles. To test this hypothesis, one could rerun the analysis in a future sensitivity test, gradually increasing the number of proper motion measurements at large distances. We leave this to future work.

\subsection{Discussion}

The underestimation of the total mass appears to be a systematic bias in our method, as it applies to these MUGS2 galaxies --- especially the regulated variety. Although it may be tempting to extrapolate results from these blind tests to our previous MW results using GCs in Paper 3, it is important to make this kind of inference cautiously, for a few reasons:

\vspace{2ex}

(1) \emph{The GC analogue populations are dissimilar to the MW, and the MUGS2 galaxies, in the end, may not be very MW-like.} The total number and number distribution of the GC analogues from the MUGS2 simulations differ substantially from that of the MW GC population (Table~\ref{tab:numberoftracers} and Figure~\ref{fig:numdensity}(b)). Additionally, the velocity profiles of the GC analogues from the regulated galaxies are dissimilar to the MW's GC velocity profile in the inner regions (Figure~\ref{fig:velprofiles}). Instead, the velocity profiles of the unregulated galaxies appear more similar to the MW, even though these galaxies lacked appropriate feedback mechanisms in the MUGS2 simulations.

\cite{keller2016} also note that the unregulated galaxies do not follow the standard stellar-mass-to-halo-mass relation. During the last stages of evolution in the MUGS2 simulations, feedback mechanisms were unable to effectively expel gas from the unregulated galaxies, which led to an overproduction of stars in their disks. Ultimately, each of these unregulated galaxies formed a massive bulge at its center which eventually depleted its gas reservoir, and created a strong central Keplerian potential \citep[see][Figure 4]{keller2016}. Given our basic assumption of a single power law for the gravitational potential, it is thus not surprising that the unregulated galaxy masses are recovered more reliably than those of the regulated galaxies (Figure~\ref{fig:mvirMWclones}).

\vspace{2ex}

(2) \emph{Only a single random sample of GC analogues from each MUGS2 galaxy was used in the analysis.} The data were subsampled because the number of GC analogues in each MUGS2 galaxy is almost always larger than the MW GC population (Table~\ref{tab:numberoftracers}). An interesting statistical test would be to repeat the analysis of Section~\ref{sec:results4} on multiple random samples from these galaxies, in order to fully understand the reliability of the mass and mass profile estimates.	However, such an investigation would be a robustness test for each galaxy and would not improve the individual estimates for each galaxy. Rather, it would test whether the credible regions have equivalent coverage probabilities. Moreover, this is computationally expensive and would only provide insight for this particular set of simulations, which may or may not accurately represent nature.

	\vspace{2ex}

(3) \emph{Our method assumes that the galaxy is in virial equilibrium and that all GC analogues are bound --- violations of these assumptions may lead to erroneous mass estimates.} The MUGS2 galaxies have complex formation histories, and consequently a mixture of bound and unbound GC analogues.   In particular, recent major mergers may create many unbound GC analogues. Unbound tracers will have higher total speeds than bound tracers; if the model assumes that unbound tracers are actually bound, then one would expect an overestimate of the mass. Indeed, this seems to be the case for $g21647$ and $g15807$.

Galaxy $g21647$ had a very recent merger event (remnants of which are visible in the mock image of the galaxy, Figure~\ref{fig:galimages}), assembling half of its mass at $z=0.2$. Unregulated galaxy $g15807$ had its last major merger at $z=2.5$, and a feature visible in velocity space indicates that it has not fully recovered from this interaction (Figure~\ref{fig:velprofiles}). Thus, these recent interactions could explain the overestimates of the galaxies' masses.

However, there is contradictory evidence for the other galaxies. On average, the regulated galaxies have a higher percentage of unbound GC analogue tracers than the unregulated galaxies (Figure~\ref{fig:mergerhistory}), and yet almost all of the regulated galaxies are underestimated (Figure~\ref{fig:mvirMWclones}). This is unexpected; assuming tracers are bound when they are not should lead to an overestimate of the mass. One possible explanation could be the location of complete data, discussed next.

(4) \emph{The mock observations of GC analogues from the regulated galaxies and unregulated galaxies differ in their completeness.} The regulated galaxies have a limited number of GC analogues within the inner regions of the galaxy (Figure~\ref{fig:velprofiles}).  Thus, when mock observations were created, most of the regulated galaxies' inner GCs analogues were given complete data within 20kpc (Section~\ref{sec:createincomplete} and e.g. Figure~\ref{fig:eprofiles}). Moreover, since mock observations beyond 20kpc were made mostly incomplete, the regulated galaxies also have a higher percentage of incomplete data than the unregulated galaxies overall (Figure~\ref{fig:mergerhistory}).

GME treats unknown proper motions as nuisance parameters in the model, and samples those nuisance parameters under the assumption that the tracers are bound to the galaxy. Thus, it is possible that the velocity estimates of the outermost incomplete data are indirectly influenced by the information from the inner tracers that have complete data. Since all GC analogues follow the same single power-law gravitational potential, this could lead to a lower mass estimate.  The complete data in the inner regions, however limited, may have enough leverage to influence the model fit when the data at large distances are incomplete. 

The evidence from our detailed investigation of regulated galaxy $g1536$ (Section~\ref{sec:energyprofiles}), coupled with the sensitivity test (Section~\ref{sec:sensitivity}), supports this hypothesis. One could test this hypothesis in future work by re-running the analysis, gradually adding more complete data at larger radii. 

\vspace{2ex}
	
(5) \emph{Some MUGS2 galaxies are exceptional cases.} Galaxies $g21647$ and $g15807$ were already mentioned as exceptional due to their merger histories. Other examples include $g22795$ and $g4720$. Both galaxies' masses were underestimated, but both galaxies are also very compact (Figure~\ref{fig:galimages}), have high concentrations of tracers at their centers (Figure~\ref{fig:velprofiles}), and have the lowest gas fractions of all eighteen MUGS2 galaxies. Moreover, the Bayesian marginal posterior distributions for $\gamma$ reached the upper limit of the prior distribution (i.e. $0.7$) in both cases, indicating the model was attempting to explore values of $\gamma>0.7$ and possibly a Keplerian potential ($\gamma\rightarrow1$). This is exceptional behaviour observed in the posterior brings us to the next important caveat to our results.

\vspace{2ex}

(6) \emph{In a Bayesian analysis, a posterior distribution that is truncated by the upper or lower limit of the prior distribution should inspire suspicion in the results.} Almost all of the regulated galaxies are inadequately matched by our model choice and prior distributions, with the $\gamma$ parameter reaching extreme ends of allowable values. In Figure~\ref{fig:jointphigam_reg}, the mode of the joint distribution for $\Phi_o$ and $\gamma$ implies the free parameter $\gamma \rightarrow 0.3$ (the lower limit of $p(\gamma)$) for $g5664$, $g8893$, $g1536$, $g422$, $g28547$, and $g24334$. These galaxies' masses were also very underestimated. Similar behaviour is seen in the posterior distributions of some unregulated galaxies too (Figure~\ref{fig:jointphigam_unreg}). Thus, in an analysis of real data, if the posterior distribution occupies extreme parts of parameter space, then any inference should be performed with caution.

The only joint posterior distributions that look reasonable are those for $g22437$, $g19195$, and maybe $g25271$ --- and these three galaxies had masses who were estimated well within the 95\% c.r.

In retrospect, using a uniform prior distribution on $\gamma$ between 0.3 and 0.7 does not necessarily reflect our prior assumptions. A value of $\gamma = 0.5$ corresponds to a Navarro-Frenk-White-like gravitational potential at large radii. A prior centered on $\gamma=0.5$ and that drops to lower probability on both sides, for example, could instead be adopted; we leave this to future work.

\vspace{2ex}

The results of the blind tests, even with these caveats, provide some insight into the behaviour of the hierarchical Bayesian method as it applies to the MUGS2 data. The main results is that an abundance of complete data in the inner regions of the galaxy, and a lack of complete data in the outer regions, might bias the total mass estimate and cumulative mass profile to lower values if the gravitational potential is assumed to follow a single power law.

	\section{Conclusions}\label{sec:conclude4}

	We have applied the hierarchical Bayesian mass estimation technique presented in Paper 3 to mock data from eighteen MUGS2 hydrodynamical MW-type galaxies \citep{keller2015, keller2016}. Our method recovered the total mass within the 95\% c.r. in 8 out of 18 cases, or 13 out of 18 cases, depending on which GC analogues were used in the analysis. 	The detailed analyses of $g15784$ and $g1536$, examples of an unregulated and a regulated galaxy, showed only moderate recovery of the cumulative mass profiles.
	
	We can cautiously say that the hierarchical method with the current model for the gravitational potential (Equation~\ref{eq:potential4}) tends to underestimate the total mass, at least for this small sample of galaxies. In particular, it is difficult for the model to predict the total mass accurately when many tracers are unbound to the galaxy and when those tracers have incomplete velocity measurements. Regardless, given the diversity of these galaxies (Figure~\ref{fig:galimages}) and our simple assumption for the gravitational potential, the method performs reasonably well in predicting the mass within the 95\% c.r.
	
	It is difficult to assess the reliability of our method on the MW data (i.e. the results of Paper 3) given the result of this study. Eighteen simulated galaxies is by no means a large sample size.  Furthermore, nine of these galaxies (the unregulated ones) are not MW-like, and the other nine galaxies (the regulated ones) do not have MW-like GC-analogue populations. The GC analogues may not be representative of a GC population in a MW-type galaxy, insofar as the MW is typical for one of its shape, size, and mass. 
	
	The detailed case of $g1536$ and the results of the sensitivity analysis also suggest that the location of complete and incomplete data in the regulated galaxies may have played a role in underestimating the total masses. Thus, not only are simulations with GC analogues that are more similar to the MW's GC population needed for future tests of the method, but the incomplete data at large radii are also a key piece of the puzzle.
	
	Based on the results of this study, we suggest modifying the use of GME in future applications to the MW: it might be prudent to use only data at large distances. The trade-off may be a more uncertain result, but with less risk of bias. The Bayesian c.r. in the resulting mass estimate and cumulative mass profile will be larger, but they will be more likely to contain the truth. Additionally, we should be cautious if the posterior distribution approaches extreme ends in the allowed parameter space. In an upcoming paper, we will apply GME to only the outer tracers of the MW, whose proper motions are available from \emph{Gaia} DR2 and the HSTPROMO project.
	
	There are a variety of avenues for future work. One way forward is to compare the viability of different galaxy model assumptions within our hierarchical framework through the Bayes factor \citep{jeffreys1939}. However, this is complicated by the shortage of analytic DFs for galaxy models. Analytic DFs are required in the current setup of our hierarchical Bayesian framework. Non-analytic models might be possible with Approximate Bayesian Computation (ABC) or ``Forward Modeling'', at the cost of substantial overhaul of the hierarchical code.
	
	However, we should not immediately discount the idea that the galaxy model employed here, although simple, may still be a good predictor of the Galaxy's mass and mass profile if we can understand how best to use it. If this is the case, then it would be a favourable alternative to computationally heavy methods like ABC for computing the mass of the MW (and in the future, other galaxies), especially with the deluge of data coming from \emph{Gaia} and LSST in the near future.

	Thus, the results of this study encourage us to pursue our investigations of simulated galaxies.  A more thorough analysis involving repeated sampling of the MUGS2 data will provide us with a better understanding of both the model choice and the method. Additionally, we plan to investigate the effects of choosing different hyperprior distributions on the model parameters, especially for $\gamma$. Furthermore, by increasing the number of complete measurements at larger radii, we will be able to investigate how well the model predicts the mass profile in the presence of more complete data at larger distances. The latter two are the most important next steps.
		
	The type of blind test performed here can also be completed with mock data from other high-performance computer simulations that produce MW-type galaxies. In particular, data from the \emph{Apostle, Aquarius, Eagle, Fire}, \emph{Illustris}, and \emph{Latte} \citep{springel2008aquarius, hopkins2014fire, vog2014illustris, schaye2015eagle,  sawala2016apostle, wetzel2016ApJ, sanderson2018} projects would all make interesting candidates.
		
	Currently, we are pursuing this avenue of research with the Modeling Star cluster population Assembly In Cosmological Simulations within the EAGLE (E-MOSAICS) project \citep{pfeffer2017}. In addition, we are investigating the importance of complete data for tracers at large radii, as well as the choice of prior distributions for the model parameters. Our findings, including results from future tests using the E-MOSAICS data --- which contain resolved GCs within a cosmological simulation --- will follow in a future paper.

	\acknowledgments
	
	This research was funded in part by an Alexander Graham Bell Canada Graduate Scholarship-Doctoral (CGS-D) to the first author, from the Natural Sciences and Engineering Research Council of Canada (held at McMaster University), and by the Moore, Sloan and Washington Research Foundations Data Science fellowship at the eScience Institute, University of Washington (UW). The lead author would like to thank Mario Juri\'c in the Department of Astronomy and DIRAC Institute and Tyler McCormick in the Department of Statistics (both at UW) for very helpful discussions regarding this research. The authors would also like to thank the anonymous referee for a thorough report that helped improve this paper.
	
	\software{R Statistical Software Environment \citep{R}, including the following packages:  \emph{stats} (part of base R), \emph{CODA:  Convergence Diagnosis and Output Analysis for MCMC} \citep{coda, codapackage},   \emph{emdbook: Ecological Models and Data in R} \citep{emdbook, emdbookBOOK}, \emph{ggplot2} \citep{hadleyGGPLOT}, \emph{MASS: Modern Applied Statistics with S} \cite{mass}, \emph{moments: Moments, cumulants, skewness, kurtosis and related tests} \citep{moments}, \emph{pracma: Practical Numerical Math Functions} \citep{pracma}, \emph{RColorBrewer: ColorBrewer Palettes}, and \emph{SNOW: Simple Network of Workstations} \citep{snow}. \citep{rcolorbrewer}. The MUGS2 galaxy simulations used GASOLINE2 \citep{wadsley2004gasoline, wadsley2017gasoline2}.}
	
	\bibliographystyle{aasjournal}
	\bibliography{myrefsP4}
	
\end{document}

%% file: MUGS2_results.tex
% latex table generated in R 3.4.4 by xtable 1.8-2 package
% Mon Jul 23 14:11:39 2018
\begin{table}[ht]
\centering
\caption{MUGS2 Galaxy Identifiers (ID) and Properties.}
\begin{tabular}{lrrrrrrrr}
  \hline
ID & $M_{200}$ & $M_*$ & $M_{\text{gas}}$ & $\lambda$' & $r_{200}$ & $r_h$ & $r_s$ & $N$ \\  
  \hline
  \hline
g7124 & 36.6 & 0.5 & 5.0 & 0.04 & 143 & 4.7 & 3.3 & 247 \\ 
  g5664 & 47.7 & 0.9 & 7.3 & 0.03 & 157 & 3.8 & 2.9 & 352 \\ 
  g8893 & 58.0 & 0.7 & 9.1 & 0.07 & 167 & 8.2 & 2.1 & 64 \\ 
  g1536 & 64.9 & 1.9 & 10.4 & 0.03 & 174 & 6.5 & 2.9 & 311 \\ 
  g21647 & 74.4 & 1.2 & 10.1 & 0.07 & 181 & 3.0 & 2.2 & 1055 \\ 
  g422 & 76.2 & 1.5 & 12.4 & 0.03 & 183 & 7.1 & 2.6 & 251 \\ 
  g3021 & 97.8 & 3.6 & 15.1 & 0.04 & 199 & 4.1 & 3.1 & 720 \\ 
  g28547 & 98.5 & 1.6 & 16.7 & 0.11 & 200 & 6.6 & 2.1 & 638 \\ 
  g24334 & 102.2 & 2.6 & 15.3 & 0.05 & 202 & 5.9 & 2.5 & 534 \vspace{0.5ex}
  \\ 
  \hline
  g22437 & 85.2 & 9.0 & 7.3 & 0.01 & 190 & 0.7 & 3.5 & 1939 \\ 
  g22795 & 85.2 & 10.6 & 4.6 & 0.01 & 190 & 0.9 & 3.1 & 1972 \\ 
  g19195 & 101.6 & 7.1 & 9.3 & 0.04 & 202 & 0.6 & 3.4 & 3041 \\ 
  g4720 & 102.5 & 14.2 & 5.5 & 0.01 & 202 & 0.5 & 3.1 & 2216 \\ 
  g4145 & 119.5 & 15.0 & 8.1 & 0.03 & 213 & 1.2 & 3.8 & 1683 \\ 
  g25271 & 125.5 & 15.6 & 7.9 & 0.02 & 216 & 1.2 & 3.3 & 2877 \\ 
  g15784 & 131.2 & 13.0 & 11.4 & 0.04 & 220 & 1.5 & 3.6 & 2381 \\ 
  g15807 & 203.2 & 21.4 & 17.5 & 0.03 & 254 & 1.0 & 2.9 & 5106 \\ 
  g27491 & 214.7 & 18.8 & 20.8 & 0.04 & 259 & 2.4 & 2.9 & 4221 \\ 
   \hline
   
\multicolumn{9}{p{\linewidth}}{NOTE: The horizontal line delineates the regulated (upper half) from unregulated (lower half) galaxies. The total mass ($M_{200}$), stellar mass ($M_{*}$), and gas mass ($M_{gas}$) for each galaxy are in units of $10^{10}$\msun, and the radius that encloses $M_{200}$ ($r_{200}$), the half-stellar mass radius ($r_h$), and the scale radius ($r_s$) are in kpc. $N$ is the total number of GC analogs in the simulation (Section~\ref{sec:GCanalogs}), and $\lambda$' is the spin parameter.}

\label{tab:numberoftracers}

\end{tabular}
\end{table}